\documentclass[aps,pre,10pt,showpacs,twocolumn,nofootinbib,floatfix]{revtex4-1}

\usepackage{graphicx,amsmath,amssymb}
\usepackage[usenames]{color}
\usepackage[latin1]{inputenc}
\usepackage{ulem}

\newcommand{\eq}[1]{Eq.~(\ref{#1})}
\newcommand{\fig}[1]{Fig.~\ref{#1}}
\newcommand{\tab}[1]{Table~\ref{#1}}
\newcommand{\sect}[1]{Section~\ref{#1}}
\newcommand{\avg}[1]{\langle #1 \rangle}
\newcommand{\olcite}[1]{Ref.~\cite{#1}}
\newcommand{\ahum}[1]{``#1''}
\newcommand{\comment}[1]{ }

\newcommand{\Ucon}{U_{1,\rm con}}
\newcommand{\Udis}{U_{1,\rm dis}}
\newcommand{\tc}{T_{\rm c}}
\newcommand{\tcv}{T_{\rm c}^{\rm VFB}}
\newcommand{\tcnb}{T_{\rm c}^{\rm NB}}
\newcommand{\hc}{H_{\rm c}}
\newcommand{\plm}{P_L(m)}
\newcommand{\plmi}{P_{L,i}(m)}
\newcommand{\calp}{{\cal P}_L(\avg{|m|})}

\newcommand{\zcp}{z_{\rm crit}}
\newcommand{\dchi}{\chi_{4}}
\newcommand{\beq}{\begin{equation}}
\newcommand{\eeq}{\end{equation}}
\newcommand{\bea}{\begin{eqnarray}}
\newcommand{\eea}{\end{eqnarray}}

\pacs{75.50.Lk,75.40.Mg,05.70.Jk,64.70.F-}

\begin{document}

\title{Finite size scaling in Ising-like systems with quenched random fields: 
Evidence of hyperscaling violation}

\author{R. L. C. Vink and T.~Fischer}
\affiliation{Institute of Theoretical Physics, Georg-August-Universit\"at 
G\"ottingen, Friedrich-Hund-Platz~1, D-37077 G\"ottingen, Germany}

\author{K. Binder}
\affiliation{Institut f\"ur Physik, Johannes Gutenberg-Universit\"at, Staudinger 
Weg 7, D-55099 Mainz, Germany}

\begin{abstract} In systems belonging to the universality class of the random 
field Ising model, the standard hyperscaling relation between critical exponents 
does not hold, but is replaced by a modified hyperscaling relation. As a result, 
standard formulations of finite size scaling near critical points break down. In 
this work, the consequences of modified hyperscaling are analyzed in detail. The 
most striking outcome is that the free energy cost $\Delta F$ of interface 
formation at the critical point is no longer a universal constant, but instead 
increases as a power law with system size, $\Delta F \propto L^\theta$, with 
$\theta$ the violation of hyperscaling critical exponent, and $L$ the linear 
extension of the system. This modified behavior facilitates a number of new 
numerical approaches that can be used to locate critical points in random field 
systems from finite size simulation data. We test and confirm the new approaches 
on two random field systems in three dimensions, namely the random field Ising 
model, and the demixing transition in the Widom-Rowlinson fluid with quenched 
obstacles. \end{abstract}

\maketitle

\section{Introduction}

Understanding the effects of quenched random disorder on phase transitions has 
been a longstanding challenge \cite{1,2,3,4,5,6}. Analysis of experiments on 
such systems is typically more difficult than work on pure systems \cite{4}. 
Theoretical methods are hampered by the fact that, for spin glasses and systems 
exposed to random fields, the marginal dimension $d^\star=6$ (the marginal 
dimension is the dimension above which mean field theory is believed to be 
reliable). In contrast, for pure systems, $d^\star=4$ \cite{8,9}. As a 
consequence, predictions of renormalization group expansions in 
$\epsilon=d^\star-d$ dimensions tend to be less reliable in the physically 
relevant dimensions $(d=2,3)$ when quenched disorder comes into play. Computer 
simulations, albeit very useful for the study of critical phenomena in pure 
systems \cite{10,11,12}, suffer from the problem that for systems exhibiting 
quenched random disorder an additional average over many samples drawn from the 
distribution characterizing the disorder needs to be taken. The disorder 
average, denoted $[ \cdot ]$, comes in addition to the usual thermal average, 
denoted $\avg{\cdot}$, and hence the computational effort is order of magnitudes 
larger. Since most analysis of critical phenomena by simulations \cite{10,11,12} 
relies on finite size scaling \cite{13,14,15,16,17}, lack of self-averaging in 
random systems \cite{18,19,20,fytas} is also a problem.

For Ising ferromagnets diluted with nonmagnetic impurities there is no doubt 
that the transition, from the high-temperature disordered to the low-temperature 
ordered phase, remains second order in $d=2,3$ \cite{3}. In addition, the 
hyperscaling relation \cite{8} between critical exponents remains valid
\begin{equation}\label{eq1} %% standard hyperscaling
 2 - \alpha = 2 \beta + \gamma = d \nu \quad (d=2,3), 
\end{equation}
and rather accurate estimates for these exponents are available \cite{22} (we 
use standard symbols to denote the exponents; definitions are provided in 
\sect{sec:theory}). For Ising ferromagnets in random fields, however, the 
situation is radically different. In $d=3$ dimensions, the existence of a 
transition at nonzero temperature was controversial until a proof for the 
existence of a spontaneous magnetization settled this issue \cite{23}; rigorous 
results on the order of the phase transition in the $d=3$ random field Ising 
model (RFIM) are still lacking however. If one accepts the evidence from 
numerical studies \cite{24, 25} that the transition is second order, it must 
have very unconventional critical behavior \cite{26,27,28}. The key point is 
that the standard hyperscaling relation, \eq{eq1}, no longer holds, but is 
replaced by
\begin{equation}\label{eq2} %% modified hyperscaling
 2 - \alpha = 2 \beta + \gamma = \nu (d-\theta).
\end{equation}
The exponent $\theta$, called the \ahum{violation of hyperscaling exponent}, is 
believed to be \cite{29,30,31}
\begin{equation}\label{eq3}
 \theta = \gamma / \nu = 2-\eta,
\end{equation}
where the critical exponent $\eta$ describes the decay of the spin pair 
correlation function right at the critical temperature \cite{8}. In addition, it 
is believed \cite{26,27,28} that critical slowing down in the RFIM is not 
described by the usual power law for the relaxation time $\tau \propto \xi^z$, 
with $z$ the \ahum{dynamic critical exponent} \cite{32} and $\xi$ the 
correlation length, but instead is governed by a much more severe 
\ahum{thermally activated critical slowing down} \cite{26,27,28}
\begin{equation}\label{eq5}
 \ln \tau \propto \xi^\theta,
\end{equation} 
with $T \to \tc$ from above. It should not come as a surprise that 
Eqs.~(\ref{eq2}),~(\ref{eq3}) and~(\ref{eq5}) make the study of the RFIM by 
Monte Carlo (MC) methods very difficult, and early studies even claimed a weak 
first order transition \cite{33}. Another problem is that standard finite size 
scaling formulations typically rely on the validity of hyperscaling, which does 
not hold in the RFIM \cite{10, 15}. Some of the consequences resulting from the 
violation of hyperscaling, \eq{eq2}, were already noted in previous works 
\cite{34}, and exploited in recent studies of colloid-polymer demixing in random 
porous media \cite{6}.

The purpose of the present paper is to analyze the consequences of \eq{eq2} for 
finite size scaling in more detail. In particular, we shall focus on the free 
energy barrier $\Delta F_L$ separating the coexisting phases for $T \leq \tc$ in 
a finite system of linear extension $L$. It has been found that $\Delta F_L$ 
increases quite strongly with $L$ at $T=\tc$ in the RFIM \cite{fyt2}. This 
behavior is puzzling because a growing barrier is usually associated with a 
first-order phase transition. In this work, the theoretical justification for 
this behavior is provided. We will show that the barrier, which in the regime 
where the transition is first-order scales as
\begin{equation}\label{eq6}
 \Delta F_L = 2 f_{\rm int} L^{d-1} \quad (T<\tc),
\end{equation}
with $f_{\rm int}$ the interfacial tension \cite{16}, right at the critical 
point is related to the hyperscaling violation critical exponent
\begin{equation}\label{eq7}
 \Delta F_L \propto L^\theta \quad (T=\tc).
\end{equation}
The factor-of-two in \eq{eq6} is a consequence of periodic boundary conditions, 
which induce two interfaces in the system when $T<\tc$. We will provide 
numerical evidence in favor of \eq{eq7} using simulation results obtained for 
two random field systems in $d=3$ dimensions, namely the RFIM and the 
Widom-Rowlinson fluid with quenched obstacles. We emphasize that in $d=2$ 
dimensions the RFIM is without a phase transition, in which case the analysis of 
the present paper does not apply.

\section{Theoretical Background}
\label{sec:theory}

We consider a system of $N$ Ising spins, situated on a $d$-dimensional lattice 
of linear extension $L$ with periodic boundaries, inside an external magnetic 
field $H$. The instantaneous magnetization per spin is defined as
\begin{equation}\label{eq:in}
 m = \frac{1}{N} \sum_{i=1}^N S_i, \quad N=L^d,
\end{equation}
with $S_i = \pm 1$ the value of the spin at the $i$-th lattice site. We assume 
that the system, in the thermodynamic limit $L \to \infty$, undergoes a 
second-order phase transition at critical temperature $T=\tc$ and field 
$H=H_{\rm c}$. Following standard practice, we introduce the relative deviations
\begin{equation}
 t \equiv T/\tc-1, \quad h \equiv H/H_{\rm c} - 1.
\end{equation}
In the vicinity of the critical point $(t=0,h=0)$, the specific heat $C$, 
susceptibility $\chi$, and correlation length $\xi$ diverge as power laws
\begin{equation}\label{eq:pl}
\begin{split}
 C(t,h=0) & \propto |t|^{-\alpha}, \quad
 \chi(t,h=0) \propto |t|^{-\gamma}, \\
 & \xi(t,h=0) \propto |t|^{-\nu}.
\end{split}
\end{equation}
In the ordered phase $T<\tc$, a finite magnetization $M$ (order parameter) 
and interfacial tension $f_{\rm int}$ develop
\begin{equation}\label{eq:op}
\begin{split}
 M(t,h=0) \propto |t|^{\beta} \quad & (t<0), \\
 f_{\rm int}(t,h=0) \propto |t|^\mu \quad & (t<0).
\end{split}
\end{equation}

We first give a heuristic derivation of the standard hyperscaling relation, 
\eq{eq1}, which is valid in pure systems, i.e.~without quenched random fields. 
Following the static scaling hypothesis \cite{36}, the singular part of the free 
energy density takes the form
\begin{equation}\label{fsing}
 f_{\rm sing}(t,h) = |t|^{2-\alpha} \tilde{f} ( h/|t|^{\beta+\gamma} ),
\end{equation}
with $\tilde{f}(x)$ a scaling function. The order parameter is obtained by 
differentiating $f_{\rm sing}$ once with respect to the external field
\begin{equation}
 M(t,h=0) \propto 
 \left. \frac{ \partial f_{\rm sing} }{ \partial h } \right|_{h=0}
 \propto |t|^{2-\alpha-\beta-\gamma},
\end{equation}
which, upon comparing to \eq{eq:op}, immediately yields $2 - \alpha = 2\beta + 
\gamma$ \cite{36}. Near criticality, the singular part of the free energy can be 
attributed to correlated regions of spins (clusters) of linear dimension $\xi$ 
\cite{36,8}. Each cluster has essentially one Ising degree of freedom 
(magnetization direction up or down), and can orient independently from its 
neighbors. Thus, while at $T \to \infty$ the total free energy $F$ of the system 
is due to the entropy of $N$ non-interacting spins, $F=- (k_B T \ln 2) N$, near 
$\tc$ we can attribute the singular part of $F$ to the entropy of $N/\xi^d$ 
independent clusters of spins, $F = -(k_B T \ln 2) N/\xi^d$, and hence
\begin{equation}\label{eq8}
 f_{\rm sing}(t,0) \propto \xi^{-d} \propto |t|^{d\nu},
\end{equation}
where in the last step \eq{eq:pl} was used. Comparing the above equation to 
\eq{fsing}, the standard hyperscaling relation $2-\alpha = d \nu$ immediately 
follows.

For an Ising system in quenched random fields near criticality the situation is 
different. To be specific, consider a random field $\pm r$ acting on each spin 
(with the signs $\pm$ drawn with equal probability such that $[r]=0$). We can 
still split the system into clusters of linear dimension $\xi$, such that each 
cluster may be considered as independent from its neighbors. However, the main 
contribution to $f_{\rm sing}$ in this case is not the entropy, \eq{eq8}, but 
rather the Zeeman energy due to the coupling to the random field. In a region of 
volume $\xi^d$, the sum of the random fields exhibits Poissonian fluctuations 
$\pm r \xi^{d/2}$. The random field excess per spin is therefore of order
\begin{equation}\label{eq:dr}
 \Delta_R \propto \pm r \xi^{-d/2},
\end{equation}
which may be conceived as an external field acting on the spins in the region. 
This implies a finite magnetization per spin
\begin{equation}\label{eq10}
 \avg{m} \propto \chi \Delta_R \propto \pm r \chi \xi^{-d/2},
\end{equation}
with $\chi$ the susceptibility. The Zeeman contribution to the free energy 
thus becomes
\begin{equation}\label{eq11}
 f_{\rm sing}(t,0) \propto \avg{m} \Delta_R \propto r^2 \chi \xi^{-d} 
 \propto |t|^{d\nu - \gamma},
\end{equation}
which dominates the entropy contribution, \eq{eq8}, upon approach of the 
critical point $t \to 0$. If we insist that $f_{\rm sing}$ retains the scaling 
form of \eq{fsing}, it follows that $2-\alpha=d\nu-\gamma$; using 
$\theta=\gamma/\nu$ (\eq{eq3}) then yields the modified hyperscaling relation 
(\eq{eq2}).

Next, we consider the exponent $\mu$ of the interfacial tension (\eq{eq:op}). 
Since \eq{fsing} is a free energy per volume, and since near $\tc$ the 
correlation length is the only relevant length scale, a simple dimensional 
argument implies that
\begin{equation}\label{eq13}
 f_{\rm int} \propto f_{\rm sing} \xi \to \mu = 2 - \alpha - \nu.
\end{equation}
In case of hyperscaling this implies $\mu = (d-1) \nu$. The important point of 
the present discussion is that this relation does not hold in the RFIM, since 
the hyperscaling relation, \eq{eq1}, is violated and replaced by the modified 
relation, \eq{eq2}. We can still infer that \eq{eq13} should hold, but now one 
must use \eq{eq2}, which leads to $\mu_{\rm RFIM} = (d-1-\theta)\nu$. Finally, 
we discuss fluctuations, which are typically large near phase transitions. In 
the pure Ising model, the thermally averaged magnetization plays the role of 
order parameter $M$, while the susceptibility $\chi$ reflects its thermal 
fluctuations
\begin{equation}\label{eq:sus}
 \text{pure Ising model} \to \begin{cases} 
 M = \avg{|m|}, \\
 \chi = L^d \left( \avg{m^2} - \avg{|m|}^2 \right).
 \end{cases}
\end{equation}
For the RFIM, the obvious generalizations are
\begin{equation}\label{eq17}
 \text{RFIM} \to \begin{cases} 
 M = [\avg{|m|}], \\
 \chi_{\rm con} = L^d \left[ \avg{m^2} - \avg{|m|}^2 \right],
 \end{cases}
\end{equation}
with $[\cdot]$ the disorder average (factors of $k_B T$ have been dropped in our 
definitions). $\chi$ and $\chi_{\rm con}$ are called \ahum{connected} 
susceptibilities: they reflect thermal fluctuations, which are present in both 
models, and diverge at criticality with exponent $\gamma$ (\eq{eq:pl}). Note 
that our definitions of the order parameter and susceptibilities use the 
absolute value of the instantaneous magnetization, as is commonly done in 
simulations \cite{orkoulas}.

In the RFIM, we can also define a \ahum{disconnected} susceptibility \cite{28}
\begin{equation}\label{eq:chidis}
 \text{RFIM} \to
 \chi_{\rm dis} \equiv L^d \left( [ \avg{|m|}^2 ] - [\avg{|m|}]^2 \right),
\end{equation}
which does not have its analogue in the pure model (removing the disorder 
average $[\cdot]$ trivially yields $\chi_{\rm dis}=0$). The motivation to 
introduce $\chi_{\rm dis}$ stems from the observation that $\avg{|m|}$ depends 
on the random field sample. Hence, in the disorder average, there will be 
sample-to-sample fluctuations in $\avg{|m|}$, which is precisely what $\chi_{\rm 
dis}$ corresponds to. Upon approach of the critical point, the disconnected 
susceptibility also diverges
\begin{equation}
 \chi_{\rm dis} \propto |t|^{\bar\gamma},
\end{equation}
defining a new critical exponent $\bar\gamma$. It is predicted that $\bar\gamma 
= 2\gamma$ \cite{26,27,28,29,30,31}, implying that sample-to-sample fluctuations 
dominate over thermal ones at criticality. If we substitute, in \eq{eq2}, 
$\theta \to \gamma/\nu$ and $\gamma \to \bar\gamma/2$, the modified hyperscaling 
relation becomes
\begin{equation}\label{eq21}
 2 \beta + \bar\gamma = d\nu,
\end{equation}
which is just the standard hyperscaling relation, \eq{eq1}, but with $\gamma$ 
replaced by $\bar\gamma$. 

\section{Finite size scaling}

\subsection{pure Ising model}

We first consider finite-size scaling (FSS) in the pure Ising model in $d=3$ 
dimensions. The Hamiltonian is given by
\begin{equation}
 {\cal H}_{\rm Ising} = -J \sum_{\avg{i,j}} S_i S_j - H \sum_{i} S_i, \quad J>0,
\end{equation}
with $\avg{i,j}$ a sum over nearest neighbors. In what follows, the temperature 
$T$ is expressed in units of $k_B/J$, with $k_B$ the Boltzmann constant. For the 
$d=3$ Ising model on cubic periodic lattices, the critical temperature $\tc 
\approx 4.511$ \cite{orkoulas}. The critical exponents are known relatively 
precisely, although not exactly (\tab{tab1}).

%% BEGIN TABLE
\begin{table}

\caption{\label{tab1} Critical exponents of the pure Ising model and RFIM in 
$d=3$ dimensions; see \olcite{28} for a more elaborate list of results for the 
RFIM.}

\begin{ruledtabular}
\begin{tabular}{ccl}
 & pure Ising & RFIM \\ \hline

%% beta estimates
$\beta$ & 0.326 \cite{9} & 0.0--0.02 \cite{hartmann} \\
 & & 0.06 \cite{25} \\
 & & 0 \cite{24} \\

%% nu estimates
$\nu$ & 0.630 \cite{9} & 1.14 \cite{hartmann}, 1.67 \cite{hartmann} \\
 & & 1.02 \cite{25} \\
 & & 1.1 \cite{24} \\
 & & 2.25 \cite{cao} \\

%% gamma estimates
$\gamma$ & 1.240 \cite{9} & 1.9 \cite{25} \\

%% gamma_bar estimates
$\bar\gamma$ & -- & 3.4--5.0 \cite{hartmann} \\
 & -- & 2.9 \cite{25} \\

\end{tabular}
\end{ruledtabular}
\end{table}
%% END TABLE

A key quantity in the numerical study of phase transitions is the order 
parameter distribution (OPD), denoted $\plm$, and defined as the probability to 
observe the system in a state with magnetization $m$. We assume the OPD is 
normalized: $\int_{-1}^{+1} \plm \, dm=1$. The OPD depends on the system size 
$L$, and on the control parameters $T$ and $H$. In the pure Ising model, due to 
spin reversal symmetry, the critical field $H_{\rm c}=0$, and so we set the 
external field to zero. The OPD is then an even function, $P_L(-m) = \plm$, 
irrespective of $T$ and $L$.

In the ordered phase, $T<\tc$, the transition is first-order. There exists a 
spontaneous magnetization, which may be positive or negative. The OPD is a 
superposition of two Gaussians, centered around $m = \pm m_0$, with 
exponentially small finite size effects in the peak positions \cite{borgs}. The 
definition $M=\avg{|m|}$ corresponds to (half) the peak-to-peak distance. In the 
disordered phase, $T>\tc$, the OPD tends to a single Gaussian peak centered 
around $m=0$. In both cases, the system self-averages: the peak widths decay 
$\propto L^{-d/2}$, ultimately becoming sharp $\delta$-functions.

%% BEGIN FIGURE
\begin{figure}
\begin{center}
\includegraphics[width=0.95\columnwidth]{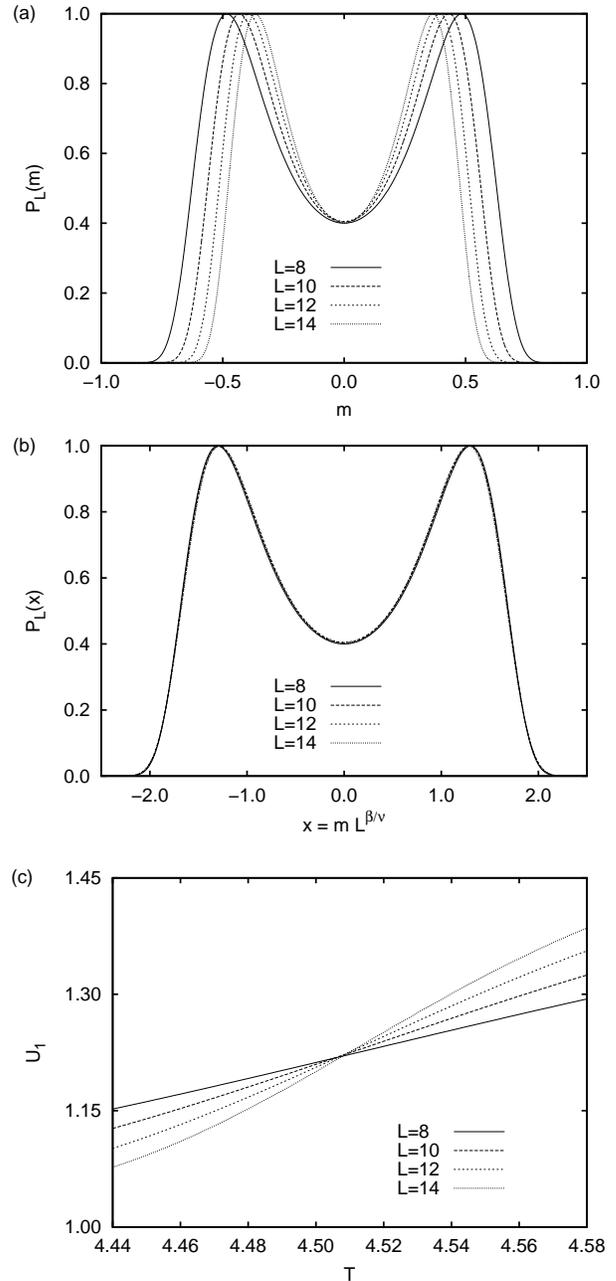}
\caption{\label{fig1} FSS in the $d=3$ pure Ising model in the critical regime. 
The linear dimension $L$ is given in units of the lattice spacing. (a) OPD 
$\plm$ obtained at $T=\tc$, $H=\hc=0$, and for several system sizes. (b) The 
same data plotted versus the scaling variable, with the critical exponents taken 
from \tab{tab1}; the data for different $L$ overlap. (c) Demonstration of the 
cumulant intersection method to locate $\tc$. Plotted is $U_1$ versus $T$ for 
several system sizes. At the critical point, the curves for different $L$ 
intersect. In the ordered (disordered) phase, $U_1 \to 1$ $(U_1\to \pi/2)$ as 
$L$ increases, in accord with \eq{eq:u1}.}
\end{center}
\end{figure}
%% END FIGURE

At criticality, the $L$-dependence of the OPD is given by the scaling 
form \cite{15, wilding}
\begin{equation}\label{eq:pmsc}
 \boxed{\plm \propto \tilde{P} (L^b m) \to 
 \avg{m^k} \propto L^{-kb} \quad (T=\tc)}
\end{equation}
with $b$ a constant, and $\tilde{P}(x)$ a scaling function characteristic of the 
Ising universality class. The standard FSS expressions for the order parameter 
and susceptibility are \cite{13,14,15}
\begin{equation}\label{eq:fss}
 M \propto L^{-\beta/\nu}, \quad 
 \chi \propto L^{\gamma/\nu} \quad (T=\tc).
\end{equation}
In order to be consistent with these expressions, \eq{eq:pmsc} requires 
$b=\beta/\nu$, and the validity of standard hyperscaling. For the pure Ising 
model, the OPD at criticality is bimodal with overlapping peaks (\fig{fig1}(a)). 
If one plots the distributions versus the scaling variable, $x = L^{\beta/\nu} 
m$, the curves for different $L$ overlap (\fig{fig1}(b)). A further consequence 
of \eq{eq:pmsc} is that cumulant ratios such as
\begin{equation}
 U_1 \equiv \avg{m^2} / \avg{|m|}^2, \quad
 U_4 \equiv \avg{m^4} / \avg{m^2}^2, \quad \rm etc.,
\end{equation}
are $L$-independent at criticality. Since the peaks in the critical OPD of the 
pure Ising model overlap (as opposed to being sharp), the corresponding cumulant 
ratios are distinctly different from the off-critical values. For example, 
considering the $U_1$ cumulant, it holds that
\begin{equation}\label{eq:u1}
 \lim_{L \to \infty} U_1 = \begin{cases}
 1 & T<\tc, \\
 U_1^\star & T=\tc, \\
 \pi/2 & T>\tc, 
 \end{cases}
\end{equation}
where $U_1^\star = 1.2391(14)$ for the $d=3$ Ising model \cite{lf}. This 
behavior is extremely useful to extract $\tc$ from simulation data 
\cite{10,11,12,15}, see \fig{fig1}(c). Note that $U_1$ is essentially the ratio 
between the order parameter and its thermal fluctuations
\begin{equation}
 \sigma_{\rm T}^2 \equiv \frac{ \avg{m^2} - \avg{|m|}^2 }{ \avg{|m|}^2 }
 = U_1 - 1.
\end{equation}
The fact that $U_1=U_1^\star$ at criticality implies that $\sigma_{\rm T}$ 
remains finite. Put differently: the thermal fluctuations of the order parameter 
$M$ remain comparable to $M$ itself at $\tc$. From this consideration one also 
understands why the peaks in the OPD are broad and overlapping. Alternatively, 
we may write
\begin{equation}\label{eq:st}
 \sigma_{\rm T}^2 = \chi / L^d M^2,
\end{equation}
from which, using \eq{eq:fss} and hyperscaling, one also immediately derives 
that $\sigma_{\rm T}$ is $L$-independent at criticality.

%% BEGIN FIGURE
\begin{figure}
\begin{center}
\includegraphics[width=0.95\columnwidth]{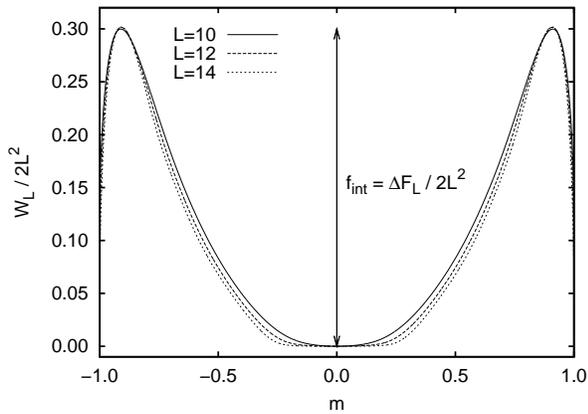}
\caption{\label{fig2} Finite size effects in the $d=3$ pure Ising model in the 
ordered phase, where the transition is first-order. Plotted is the 
scaled-and-shifted logarithm of the OPD, $W_L = \ln \plm$, for various system 
sizes at $T=3.33$, which is well below $\tc$. The peak height corresponds to the 
interfacial tension $f_{\rm int}$. Note also the flat region unfolding between 
the peaks as $L$ increases.}
\end{center}
\end{figure}
%% END FIGURE

The free energy barrier is obtained from the logarithm of the OPD $W_L = \ln 
\plm$. We define $\Delta F_L$ as the average peak height, measured from the 
minimum \ahum{in-between} the peaks. In the ordered phase, $T<\tc$, the 
transition is first-order, and the barrier is related to the interfacial tension 
$f_{\rm int}$ via \eq{eq6}. This is shown in \fig{fig2}. Note also that the peak 
positions -- at least on the scale of the graph -- do not reveal any strong $L$ 
dependence either, consistent with exponentially small finite size effects 
\cite{borgs}. The peaks also become sharper as $L$ increases, showing that the 
system is self-averaging. Finally, we note that a flat region between the peaks 
in $W_L$ unfolds as $L$ increases. This is a sign that interactions between the 
interfaces through the periodic boundaries are vanishing \cite{gm}.

%% BEGIN FIGURE
\begin{figure}
\begin{center}
\includegraphics[width=0.95\columnwidth]{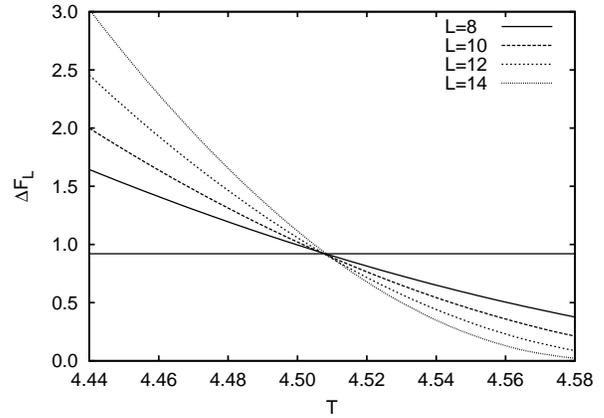}
\caption{\label{fig3} Scaling of the free energy barrier $\Delta F_L$ in the 
pure Ising model. By plotting $\Delta F_L$ versus $T$ for various $L$, the 
critical temperature appears as an intersection point. The horizontal line marks 
the (universal) value $\Delta F^\star \sim 0.9$ for the $d=3$ Ising 
model.}
\end{center}
\end{figure}
%% END FIGURE

Precisely at criticality, the scaling of the barrier is different. We may still 
assume that \eq{eq6} holds, but on a length scale that is set by the correlation 
length
\begin{equation}\label{eq:df}
 \Delta F_\xi \propto f_{\rm int} \xi^{d-1} 
 \propto \xi^{d - 1 - (2-\alpha-\nu)/\nu},
\end{equation}
where in the last step the critical power law of $f_{\rm int}$ and \eq{eq13} 
were used; by virtue of hyperscaling, the length scale drops out. Hence, the 
barrier is a constant $L$-independent value $\Delta F_L \equiv \Delta F^\star$ 
at criticality. Of course, the fact that the OPD at $T=\tc$ has a universal 
shape, see \fig{fig1}(b), also implies this property. In the pure Ising model, 
the barrier thus scales as 
\begin{equation}
 \lim_{L \to \infty} \Delta F_L = \begin{cases}
 2 f_{\rm int} L^{d-1} & T<\tc, \\
 \Delta F^\star & T=\tc, \\
 0 & T>\tc.
 \end{cases}
\end{equation}
This behavior is also well-suited to locate $\tc$ \cite{lee}. For instance, one 
plots $\Delta F_L$ versus $T$ for various $L$; at the critical point, the data 
for different system sizes intersect (\fig{fig3}).

In brief, we have summarized FSS in the pure Ising model. The important message 
is that, due to hyperscaling, the OPD assumes a universal shape at the critical 
point. As a result, the free energy barrier and selected cumulant ratios assume 
non-trivial $L$-independent values, which can be used to locate the critical 
point. We also note that, by using the intersection methods of $U_1$ and $\Delta 
F_L$ (\fig{fig1}(c) and \fig{fig3}), moderate system sizes suffice to locate 
the critical point with an accuracy better than one part in a thousand.

\subsection{RFIM}

We now consider FSS in the RFIM at its critical point. The Hamiltonian reads as
\begin{equation}\label{eq:rfim}
 {\cal H}_{\rm RFIM} = -J \sum_{<i,j>} S_i S_j - \sum_i r_i S_i 
 - H \sum_{i} S_i,
\end{equation}
$J>0$, with $r_i$ the quenched random field acting on the spin at the $i$-th 
lattice site. It is convenient to draw $r_i$ from a distribution that is 
symmetric about zero; this ensures that $\hc=0$. The most common choices are the 
bimodal distribution $P(r_i) \propto \delta(r_i - \sigma) + \delta(r_i + 
\sigma)$ and the Gaussian $P(r_i) \propto \exp(-r_i^2/2\sigma^2)$, where 
$\sigma$ is the random field strength. In contrast to the pure Ising model, the 
critical exponents of the RFIM are not known very precisely, see \tab{tab1}, 
where several exponent estimates from theoretical and simulational works are 
listed. We believe it is safe to conclude that $\beta$ is close to zero. 
Modified hyperscaling then implies $\theta=\gamma/\nu \approx 1.5$ and 
$\bar\gamma/\nu \approx 3$, where dimensionality $d=3$ is assumed.

\subsubsection{Consequences of modified hyperscaling}

One of the most striking consequences of hyperscaling violation is that the 
thermal fluctuations become negligible at the critical point. For the RFIM, the 
analogue of \eq{eq:st} becomes $\sigma_{\rm T}^2 = \chi_{\rm con} / L^d M^2$; 
using the FSS expressions $M \propto L^{-\beta/\nu}$ and $\chi_{\rm con} \propto 
L^{\gamma/\nu}$, it follows that
\begin{equation}\label{eq:decay}
 \sigma_{\rm T}^2 \propto L^{2\beta/\nu + \gamma/\nu - d}
 \propto L^{-\gamma/\nu} \quad (T=\tc),
\end{equation}
where now the modified hyperscaling relation, \eq{eq2}, was used. In the RFIM, 
$\sigma_{\rm T}$ thus decays to zero with increasing $L$, whereas in the pure 
Ising model $\sigma_{\rm T}$ saturates at a finite $L$-independent value.

Even though the thermal fluctuations vanish in the RFIM for large $L$, we must 
not forget about the sample-to-sample fluctuations, which are characterized by 
$\chi_{\rm dis}$. In line with $\sigma_{\rm T}$, we compare the order parameter 
to the magnitude of sample-to-sample fluctuations as
\begin{equation}\label{eq:sdd}
 \sigma_{\rm D}^2 \equiv \frac{ \chi_{\rm dis} }{ L^d M^2}
 \propto L^{2\beta/\nu + \bar\gamma/\nu - d} \quad (T=\tc),
\end{equation}
where also the FSS expression $\chi_{\rm dis} \propto L^{\bar\gamma/\nu}$ was 
used. The remarkable consequence of modified hyperscaling, \eq{eq21}, is 
therefore that $\sigma_{\rm D} \propto L^0$, i.e.~becoming constant at 
criticality. Hence, in the RFIM, it is the sample-to-sample fluctuations that 
\ahum{scale with $L$}, not the thermal fluctuations.

How does this modified scaling affect the OPD? First note that, in addition to 
$T$, $L$, and $H$, the probability to observe a certain instantaneous 
magnetization $m$ also depends on the random field sample. We therefore write 
$\plmi$, where the index $i$ denotes one particular sample of random fields. We 
thus have a set of distributions. In practice, this requires that $\plmi$ be 
measured for at least $i=1, \ldots ,K$ samples, where $K$ must be large enough. 
We can immediately rule out that $\plmi$ at criticality obeys the scaling form, 
\eq{eq:pmsc}, since hyperscaling is violated. Assuming that the majority of 
distributions $\plmi$ remains bimodal at $\tc$ -- which needs to be verified in 
practice -- the peak-to-peak distance scales as the order parameter $M$, while 
the squared peak widths $W^2 \propto \chi_{\rm con}/L^d$. Since $\sigma_{\rm T} 
(=W/M)$ decays to zero, see \eq{eq:decay}, it follows that the peaks in $\plmi$ 
become sharp. Again, this is in contrast to the pure Ising model, where the 
critical OPD features broad and overlapping peaks. In the RFIM, the shape of 
$\plmi$ at $T=\tc$ and $T<\tc$ is the same: bimodal with sharp non-overlapping 
peaks. The crucial difference is that, at $T=\tc$, the peak-to-peak distance 
decays $\propto L^{-\beta/\nu}$, while for $T<\tc$ the peak positions saturate 
at finite values $\pm m_0$. In the disordered phase, $T>\tc$, $\plmi$ should 
again be single peaked. As a consequence, ratios of {\it connected} 
quenched-averaged moments, such as $[\avg{m^{2k}}] / [\avg{m^k}^2]$ or 
$[\avg{m^{2k}}] / [\avg{m^k}]^2$, no longer assume \ahum{special} values at 
criticality, but equal those of the ordered phase $T<\tc$. For instance:
\begin{equation}\label{eq:u1rf}
 \lim_{L \to \infty}
 \Ucon \equiv \frac{ [\avg{m^2}] }{ [\avg{|m|}^2] }
 = \begin{cases}
 1 & T<\tc, \\
 1 & T=\tc, \\
 \pi/2 & T>\tc,
 \end{cases}
\end{equation}
which does not lend itself well to extract $\tc$ from finite-size simulation 
data.

Does this imply there is no \ahum{scaling} at all in the RFIM at its critical 
point? The answer to this question is an unequivocal \ahum{No}! Scale invariant 
distributions and observables still exist in the RFIM, but they must be 
constructed keeping modified hyperscaling, \eq{eq21}, in mind. For instance, to 
each random field sample $i$ there corresponds a distribution $\plmi$, from 
which an average magnetization $\avg{|m|}_i$ can be obtained. Due to 
sample-to-sample fluctuations, the values $\avg{|m|}_i$ will generally differ. 
Hence, it is useful to consider the distribution $\calp$, defined as the 
probability of a particular random field sample yielding a thermally averaged 
magnetization $\avg{|m|}$. In the absence of quenched disorder, $\calp$ reduces 
to a $\delta$-function; in the presence of quenched disorder, $\calp$ may retain 
a finite width. The moments of $\calp$ correspond to $[\avg{|m|}^k]$, which are 
precisely the quantities needed to compute the order parameter and the 
disconnected susceptibility. If we compare the average of $\calp$ to its 
root-mean-square width, we recover $\sigma_{\rm D}$ of \eq{eq:sdd}; by virtue of 
modified hyperscaling, the latter becomes constant at criticality. Hence, in the 
RFIM, it is the distribution $\calp$ that remains broad at criticality. Our 
\ahum{Ansatz} is therefore that the scaling of the OPD in the pure Ising model, 
is replaced by scaling of $\calp$ in the RFIM. We thus propose
\begin{equation}\label{eq:prf}
 \boxed{\begin{split}
 {\cal P}_L( \avg{|m|} ) &\propto \tilde{\cal P}(L^b \avg{|m|}) \to \\
 [\avg{|m|}^k] &\propto L^{-kb} \quad (T=\tc, \, \rm RFIM)
 \end{split}}
\end{equation}
as the analogue of \eq{eq:pmsc}, with $\tilde{\cal P}(x)$ a scaling function 
characteristic of the RFIM. Consistency of \eq{eq:prf} with $M \propto 
L^{-\beta/\nu}$ and $\chi_{\rm dis} \propto L^{\bar\gamma/\nu}$ requires 
$b=\beta/\nu$, and the validity of modified hyperscaling, \eq{eq21}. Note that 
\eq{eq:prf} also implies that \ahum{disconnected} cumulants such as
\begin{equation}\label{eq:u1dis}
 \Udis \equiv [ \avg{|m|}^2 ] / [ \avg{|m|} ]^2,
\end{equation}
become $L$-independent at $\tc$. This property suggests that a generalization of 
the \ahum{cumulant intersection method}, \fig{fig1}(c), is also feasible in the 
RFIM. In this case, one should plot $\Udis$ versus $T$; curves for different $L$ 
should intersect at $\tc$.

%% BEGIN FIGURE
\begin{figure}
\begin{center}
\includegraphics[width=0.75\columnwidth]{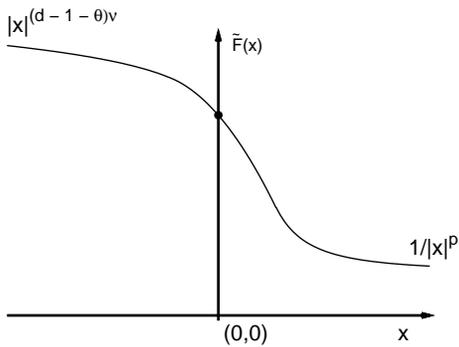}
\caption{\label{fig4} Schematic plot of the scaling function $\tilde{F}(x)$, 
defined by \eq{eq:gx}, describing the free energy barrier $\Delta F_L$ in the 
RFIM versus the scaling variable $x=t L^{1/\nu}$. There occurs a smooth 
crossover from $\tilde{F}(x) \propto |x|^{ (d-1-\theta) \nu }$ for $x \ll 0$, to 
$\tilde{F}(x) \propto 1/|x|^p$ for $x \gg 0$ (with $p>\nu\theta$).}
\end{center}
\end{figure}
%% END FIGURE

The second main consequence of modified hyperscaling concerns the scaling of the 
free energy barrier. The barrier is no longer constant at $\tc$, but rather 
$\Delta F_L \propto L^\theta$, with $\theta=\gamma/\nu$ the \ahum{violation of 
hyperscaling} exponent. This follows immediately from \eq{eq:df}, where now the 
modified hyperscaling relation must be used, as well as the FSS \ahum{Ansatz} 
$\xi \propto L$ \cite{13,14,15}. We thus expect, for random field Ising 
universality,
\begin{equation}
 \lim_{L \to \infty} \Delta F_L \propto \begin{cases}
 f_{\rm int} L^{d-1} & T<\tc, \\
 L^\theta & T=\tc, \\
 0 & T>\tc.
 \end{cases}
\end{equation}
Following standard FSS practice \cite{13,14,15}, we may also write
\begin{equation}\label{eq:gx}
 \Delta F_L = L^\theta \tilde{F}(x), \quad x=t L^{1/\nu}, 
\end{equation}
with $\tilde{F}(x)$ a scaling function. The scaling of the barrier in the 
ordered phase, $T<\tc$, implies that $\tilde{F}(x) \propto |x|^{ (d-1-\theta) 
\nu } \quad (x \ll 0)$. Precisely at criticality $x=0$, we should recover 
\eq{eq7}, i.e.~$\tilde{F}(0)>0$, while in the disordered phase $T>\tc$ the 
barrier should vanish $\tilde{F}(x) \propto 1/|x|^p, \quad p>\nu\theta \quad (x 
\gg 0)$. From these considerations, as well as from the fact that the scaling 
function must be smooth, we derive the sketch shown in \fig{fig4}. The fact that 
$\tilde{F}(x)$ is a smooth function, implies that (huge) free energy barriers 
$\Delta F_L \propto \xi^\theta$ persist above $\tc$ also (in sharp contrast to 
the pure Ising model). The latter give rise to the Arrhenius law for the 
relaxation time, \eq{eq5}.

\subsubsection{Practical considerations: tuning the external field}
\label{sec:tune}

In FSS studies, the critical region is \ahum{scanned} by varying the control 
parameters $T$ and $H$. Mathematically, this can be conceived as following a 
path in the $(T,H)$ plane. One may choose the path freely, as long as it passes 
through the critical point $(\tc,\hc)$ in the thermodynamic limit. In the pure 
Ising model and RFIM, the critical field $\hc=0$, and so the critical region may 
be scanned by varying $T$ at fixed $H=0$. We call this the symmetry path $l_S : 
H=0$. It may also happen that $\hc$ is not known beforehand. This is often the 
case in fluids, where the analogue of $H$ is the chemical potential. In these 
situations, different paths must be constructed. One example is the 
\ahum{equal-weight} path \cite{eqarea}, whereby $H$ is tuned such that the peaks 
in the OPD have equal area. The field now becomes a non-trivial function of 
temperature and system size $H = f(T,L)$. As it turns out, an infinite number of 
paths can be constructed along these lines \cite{locus}. Here, we will mostly 
use the path $l_\Gamma$, whereby $H$ is tuned such that
\begin{equation}\label{eq:lgamma}
 l_\Gamma : \partial \avg{m} / \partial H \to \text{max}.
\end{equation}
Note that, when $l_\Gamma$ is used in conjunction with quenched disorder, $H$ 
not only depends on $T$ and $L$, but also on the random field sample $i$, that 
is $H = f_i(T,L)$ with $i=1, \ldots , K$. For fixed $T$ and $L$, each sample 
thus yields its own field $H_i$. A sharp transition requires that, for $T \leq 
\tc$, the variance of $H_i$ vanishes as $[H^2]-[H]^2 \propto 1/L^d$, with $[H^p] 
= (1/K) \sum_{i=1}^K H_i^p$. It is important that the variance decays with 
exponent $d$, i.e.~there should be no critical exponent involved. The field 
$H_i$ that maximizes $\Gamma$ is just chosen to cancel the random field excess 
$\Delta_R$ of \eq{eq:dr}, the square of which scales inversely with the volume. 
The behavior of the variance above $\tc$ is less relevant because here we no 
longer have phase coexistence, and so the OPD tends to a single Gaussian peak as 
$L \to \infty$. The path $l_\Gamma$, as well as the \ahum{equal-weight} path, 
then become meaningless anyway.

\section{Numerical tests for the RFIM}
\label{sec:test}

We consider the RFIM Hamiltonian of \eq{eq:rfim}, using Gaussian random fields 
with $\sigma=1.4$, for which Newman and Barkema (NB) report as critical 
temperature $\tcnb \approx 3.6$ \cite{25}. Since the distribution of random 
fields is symmetric about zero, it also holds that $\hc=0$. The implications of 
modified hyperscaling will now be verified.

\subsection{FSS using the symmetry path $l_S$}

%% BEGIN FIGURE
\begin{figure}
\begin{center}
\includegraphics[width=0.95\columnwidth]{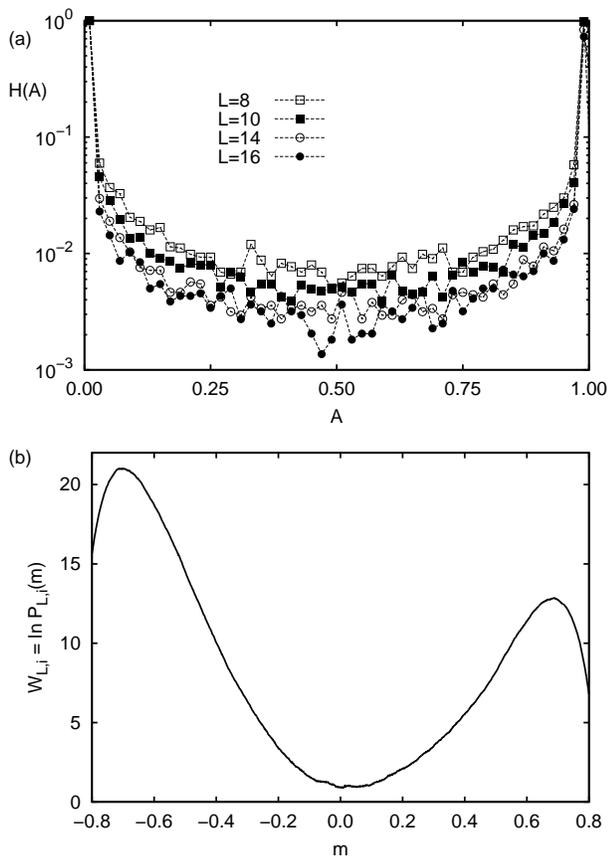}
\caption{\label{fig5} Investigation of the \ahum{typical} shape of $\plmi$ in 
the RFIM at $T=\tcnb$ using the symmetry path $l_S$. (a) Histograms $H(A)$ for 
various system sizes $L$. Note the logarithmic vertical scale. (b) Typical 
distribution $W_{L,i} = \ln \plmi$ for $L=14$; for this distribution $A\sim 1$.}
\end{center}
\end{figure}
%% END FIGURE

We first use the symmetry path. That is, we measure $\plmi$ at fixed $H=0$. Even 
though the critical field $\hc=0$, spin reversal symmetry is broken in single 
samples, and so we do not expect $\plmi$ to be symmetric (only after the 
disorder average $[\cdot]$ has been taken is the symmetry restored). To verify 
that symmetric distributions are rare, we consider the ratio $A = \int_{-1}^0 
P_{L,i} (m) \, dm / \int_{-1}^1 P_{L,i} (m) \, dm$. For a perfectly symmetric 
distribution $A=1/2$ (the reverse is not necessarily true). \fig{fig5}(a) shows 
histograms of observed $A$ values, measured at $T=\tcnb$, and for various $L$. 
For each system size, $K \sim 10,000$ random field samples were used; more 
details regarding this choice are provided in the Appendix. It is clear that 
symmetric distributions are rare. Most distributions yield a value of $A$ close 
to zero or unity, meaning that the \ahum{weight} is entirely concentrated left 
or right of $m=0$. \fig{fig5}(b) shows the logarithm $W_{L,i} = \plmi$ of one 
such \ahum{typical} distribution. A bimodal structure is revealed, but the peak 
heights are very different. If one plots $\plmi$ itself it is clear that only a 
single peak survives. We conclude: by using the symmetry path, $\plmi$ is mostly 
a single peak. However, note that $H(A=1/2)$ is not zero: distributions whose 
\ahum{weight} is spread symmetrically around $m=0$ do occasionally occur. We 
return to this point later.

%% BEGIN FIGURE
\begin{figure}
\begin{center}
\includegraphics[width=0.95\columnwidth]{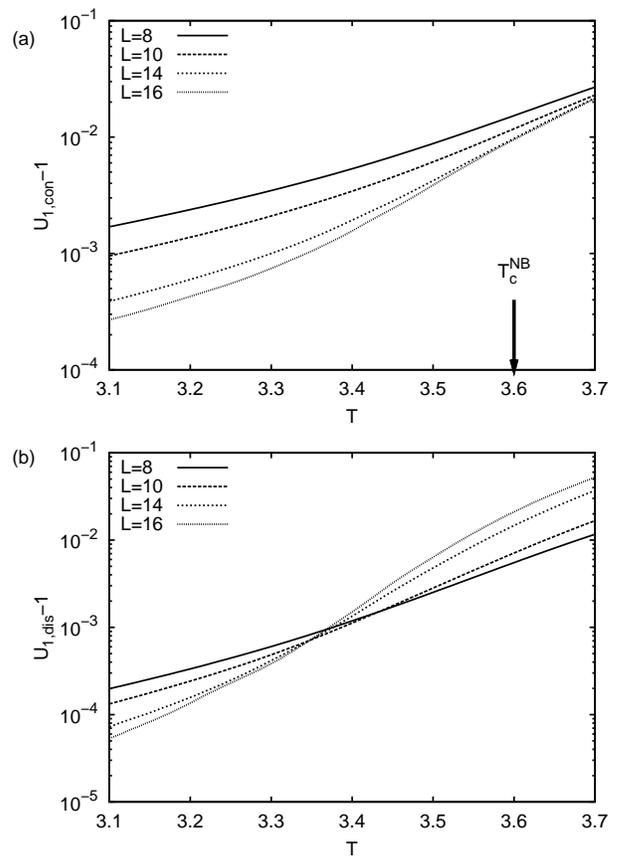}
\caption{\label{fig6} Cumulant analysis of the RFIM using the symmetry path 
$l_S$. For clarity, the curves are shifted by unity, and a logarithmic vertical 
scale is used. (a) $\Ucon$ versus $T$ for various $L$; the arrow marks 
$\tcnb$. (b) The disconnected cumulant $\Udis$ versus $T$ for various 
$L$. Note the absence of intersections in $\Ucon$, and their presence 
in $\Udis$. This is conform the modified hyperscaling scenario.}
\end{center}
\end{figure}
%% END FIGURE

The symmetry path $l_S$ does not lend itself well to measure free energy 
barriers, most distributions being single-peaked, but we can still probe 
sample-to-sample fluctuations\footnote{A \ahum{work-around} to extract the 
barrier using the symmetry path can still be defined, see 
Appendix~\ref{app:alt}.}. For each random field sample $i$, we calculate the 
magnetization $\avg{|m|}_i = \int_{-1}^{+1} |m| \plmi \, dm$, and the second 
moment $\avg{m^2}_i = \int_{-1}^{+1} m^2 \plmi \, dm$, which are then averaged 
to obtain $[ \avg{|m|} ] = (1/K) \sum_{i=1}^K \avg{|m|}_i$, and so forth. In 
\fig{fig6}(a), we plot the connected cumulant $\Ucon$ (\eq{eq:u1rf}) versus $T$ 
for various system sizes, while (b) shows the disconnected cumulant $\Udis$, 
\eq{eq:u1dis}. The striking result is that $\Udis$ reveals an intersection 
point, while $U_{1,\rm RFIM}$ does not: exactly what is predicted by modified 
hyperscaling! From the intersections in $\Udis$, we conclude that the critical 
temperature is somewhat below $\tcnb$. Above $\tc$, the connected cumulant 
$\Ucon \to \pi/2$ as $L \to \infty$. If one plots $\Ucon$ versus $T$ for two 
values of $L$, an intersection will also be found, at some value $T_L>\tc$; see 
for instance the curves for $L=14$ and $L=16$ in \fig{fig6}(a). As $L$ 
increases, $T_L$ will shift toward $\tc$, but there is no intersection of 
$\Ucon$ {\it at} $\tc$.

%% BEGIN FIGURE
\begin{figure}
\begin{center}
\includegraphics[width=0.95\columnwidth]{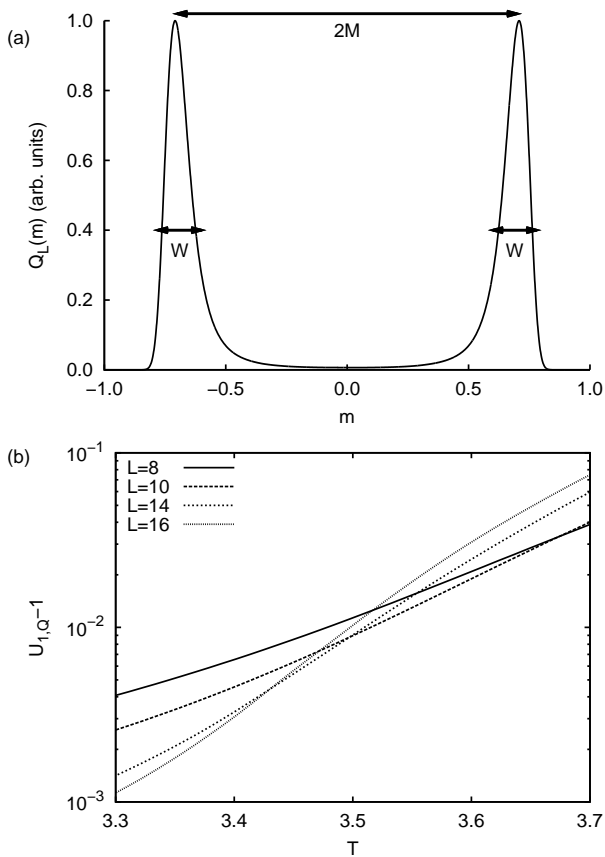}
\caption{\label{fig8} (a) The quenched-averaged distribution $Q_L(m)$ of the 
RFIM obtained at $T=\tcnb$, $L=14$, and using the symmetry path $l_S$. The 
peak-to-peak distance is proportional to the order parameter $M$, while the peak 
widths $W$ reflect the sum of thermal and sample-to-sample fluctuations. (b) The 
leading cumulant of $Q_L(m)$ versus $T$ for various $L$.}
\end{center}
\end{figure}
%% END FIGURE

We still discuss the quenched-averaged distribution $Q_L(m) \equiv (1/K) 
\sum_{i=1}^K P_{L,i} (m)$, i.e.~the arithmetic mean of the individual 
(normalized) OPDs. Since $\plmi$ is mostly a single peak, located with equal 
probability at positive or negative values, $Q_L(m)$ is bimodal and symmetric 
about $m=0$ (\fig{fig8}(a)). The peak-to-peak distance corresponds to (twice) 
the order parameter $M = [\avg{|m|}]$, but care is needed to interpret the peak 
widths $W$. The moments of $Q_L(m)$ are of the form $[\avg{m^k}]$, and so the 
peak widths correspond to
\begin{equation}
 W^2 = [ \avg{m^2} ] - [ \avg{|m|} ]^2
 = \chi_{\rm con}/L^d + \chi_{\rm dis}/L^d,
\end{equation}
which is the sum of thermal fluctuations (set by $\chi_{\rm con}$) and 
sample-to-sample fluctuations (set by $\chi_{\rm dis}$). Consequently, the 
leading cumulant of $Q_L(m)$ becomes
\begin{equation}\label{eq:u1q}
 U_{1,Q} = \frac{ [ \avg{m^2} ] }{ [ \avg{|m|} ]^2  } = 
 \Udis + \frac{ \chi_{\rm con} }{ L^d M^2 }.
\end{equation}
Using now the FSS expressions $M \propto L^{-\beta/\nu}$, $\chi_{\rm con} 
\propto L^{\gamma/\nu}$, and modified hyperscaling, we obtain
\begin{equation}
 U_{1,Q} -  \Udis \propto L^{-\gamma/\nu} \quad (T=\tc).
\end{equation}
Hence, in the thermodynamic limit, $U_{1,Q}$ becomes identical to $U_{1,\rm 
dis}$. Plotting $U_{1,Q}$ versus $T$ for different $L$ one therefore also 
observes intersections (\fig{fig8}(b)). Note that, due to the correction term 
induced by the connected susceptibility in \eq{eq:u1q}, one expects that for 
small $L$ the intersections are more scattered than those for $\Udis$; the data 
in \fig{fig6}(b) and \fig{fig8}(b) are compatible with this expectation.

\subsection{FSS using the path $l_\Gamma$}

%% BEGIN FIGURE
\begin{figure}
\begin{center}
\includegraphics[width=0.95\columnwidth]{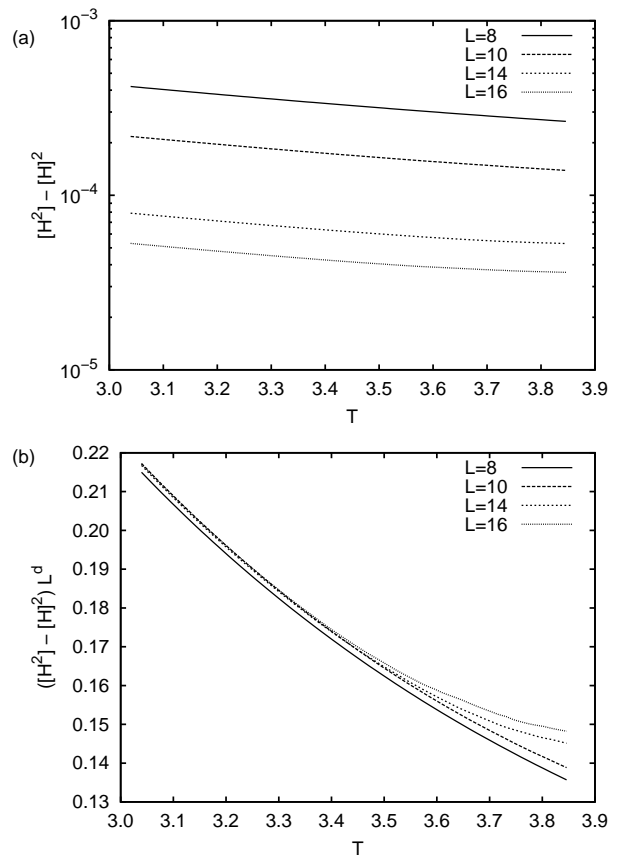}
\caption{\label{fig12} (a) Variance $[H^2]-[H]^2$ of the \ahum{tuned} external 
fields $H_i$ versus $T$ using the path $l_\Gamma$ for several system sizes. (b) 
The variance multiplied by $L^d$.}
\end{center}
\end{figure}
%% END FIGURE

%% BEGIN FIGURE
\begin{figure}
\begin{center}
\includegraphics[width=0.95\columnwidth]{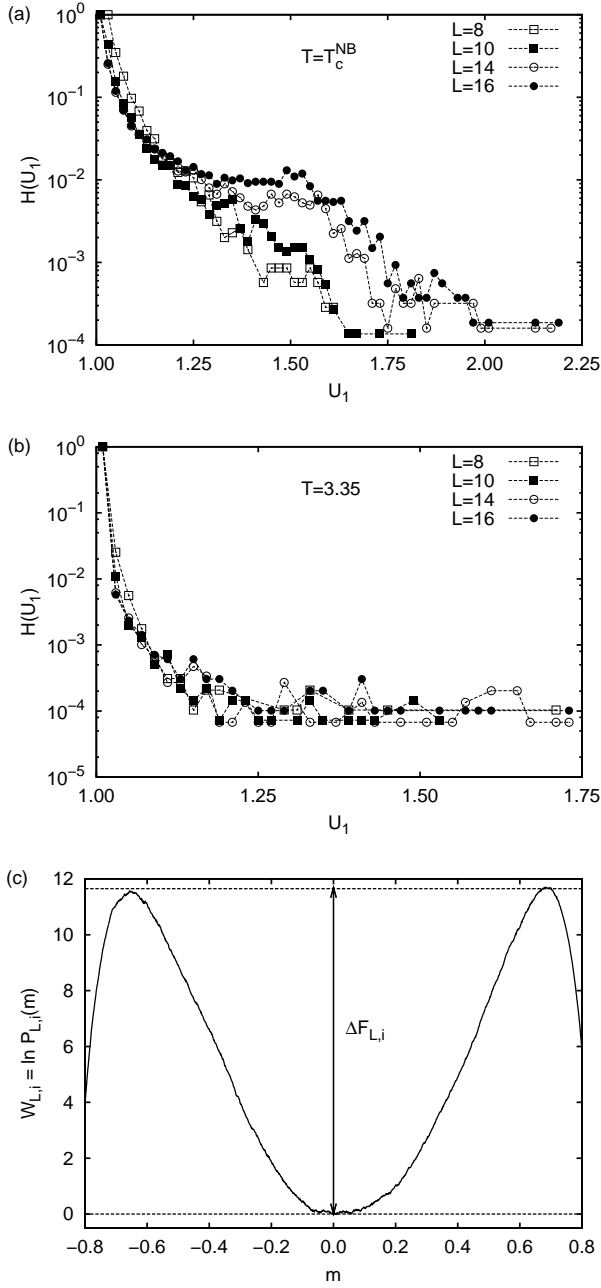}
\caption{\label{fig9} Investigation of the \ahum{typical} shape of $\plmi$ in 
the RFIM using the path $l_\Gamma$. Shown are histograms $H(U_1)$ for various 
$L$ obtained at (a) $T=\tcnb$ and (b) $T=3.35$. The histograms peak at $U_1=1$ 
implying that most distributions are bimodal. Note the logarithmic vertical 
scales. (c) Typical distribution $W_{L,i} = \ln \plmi$ obtained for $L=14$ and 
$T=\tcnb$. Since $W_{L,i}$ is bimodal, a free energy barrier $\Delta F_{L,i}$ 
can be extracted (vertical arrow).}
\end{center}
\end{figure}
%% END FIGURE

We now use the path $l_\Gamma$, where for each random field sample $i$ the 
external field $H_i$ is tuned according to \eq{eq:lgamma}. We first verify, in 
\fig{fig12}, that the variance of $H_i$ indeed decays $\propto 1/L^d$. The raw 
data are shown in (a), while (b) shows the same data multiplied by $L^d$. In the 
latter representation, the $L$-dependence should cancel for $T \leq \tc$. This 
is confirmed by the collapse of the data of the larger systems; only the $L=8$ 
data is somewhat off, which indicates that this system may be too small for an 
accurate FSS analysis. The \ahum{swaying-out} of the curves at high $T$ is a 
sign of entering the one-phase region, where the path $l_\Gamma$ becomes 
ill-defined.

By using $l_\Gamma$, we expect that most distributions become bimodal for $T 
\leq \tc$. In \fig{fig9}, we show histograms of observed cumulant values, for 
$T=\tcnb$ (a) and $T=3.35$ (b). We believe the latter temperature is closer to 
the true $\tc$, based on the intersections of the disconnected cumulant, 
\fig{fig6}(b). The histograms peak at $U_1=1$, confirming that bimodal 
distributions dominate. An example distribution is shown in \fig{fig9}(c), from 
which a barrier $\Delta F_{L,i}$ can be accurately extracted (vertical arrow). 
We remind the reader that the barrier is to be obtained from the logarithm of 
$\plmi$. Note also an important finite size effect in the histograms $H(U_1)$. 
At $T=\tcnb$, for increasing $L$, a shoulder develops at $U_1 \sim \pi/2$, 
meaning that single-peaked distributions become more likely in larger systems. 
In contrast, $H(U_1)$ at $T=3.35$ reveals no such effect. We believe this 
indicates that $T=\tcnb$ is actually above the critical temperature. Above 
$\tc$, in the thermodynamic limit, the OPD is single-peaked. Hence, 
\fig{fig9}(a) shows the evolution toward this shape. The convergence with $L$ is 
clearly very slow, and much larger systems are required before single-peaked 
distributions would dominate bimodal ones in finite-size simulation data.

%% BEGIN FIGURE
\begin{figure}
\begin{center}
\includegraphics[width=0.95\columnwidth]{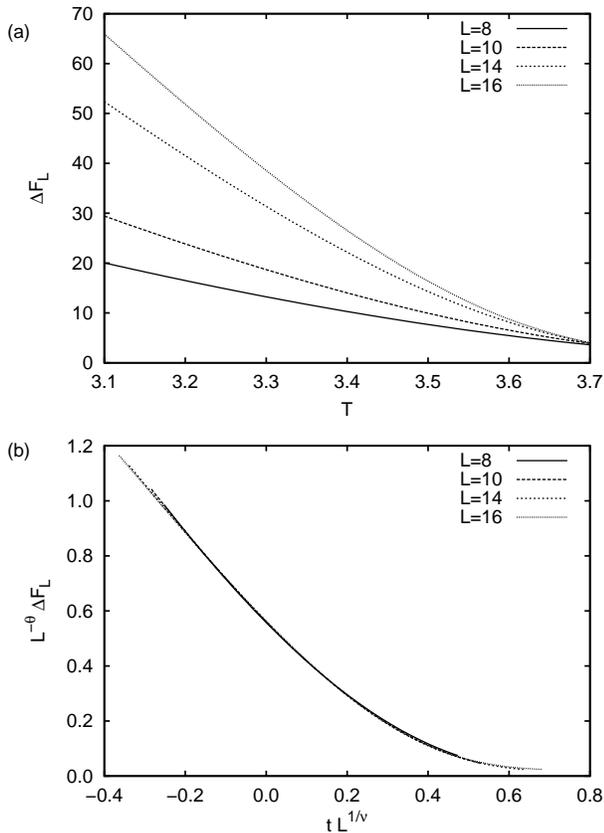}
\caption{\label{fig10} FSS of the free energy barrier $\Delta F_L$ in the RFIM 
using the path $l_\Gamma$. (a) $\Delta F_L$ versus $T$ for various system sizes. 
(b) The same data as in (a) but using scaled variables. The validity of the 
scaling form, \eq{eq:gx}, is confirmed by the collapse of the data 
from the various system sizes onto a single curve.}
\end{center}
\end{figure}
%% END FIGURE

The path $l_\Gamma$ facilitates a first test of the scaling of the 
quenched-averaged barrier $\Delta F_L = (1/K) \sum_{i=1}^K \Delta F_{L,i}$, 
where the sum is over all $K$ considered random field samples. For distributions 
where a barrier cannot be meaningfully defined, such as single or triple peaks, 
$\Delta F_{L,i}$ is set to zero. In \fig{fig10}(a), we show $\Delta F_L$ versus 
$T$ for various $L$. Following modified hyperscaling, we expect $\Delta F_L$ to 
scale conform \eq{eq:gx}. Hence, plotting $L^{-\theta} \Delta F_L$ versus $t 
L^{1/\nu}$, $t = T/\tc-1$, the curves for different $L$ should collapse, 
provided suitable values of $\theta$, $\nu$, and $\tc$ are used. This result is 
shown in \fig{fig10}(b). Here, $\theta=1.5$ was assumed, and by varying $\nu$ 
and $\tc$, a data collapse is indeed obtained (the plot uses $\nu=1.9$ and 
$\tc=3.32$). We have verified that by using $\theta=0$, i.e.~the value of the 
pure model, no data collapse is obtained. The estimate of $\nu$ is rather large, 
but still within the range of values reported in \tab{tab1}. Note also that 
$\tc$ used in \fig{fig10}(b) agrees with that of the disconnected cumulant 
intersections, \fig{fig6}(b).

%% BEGIN FIGURE
\begin{figure}
\begin{center}
\includegraphics[width=0.95\columnwidth]{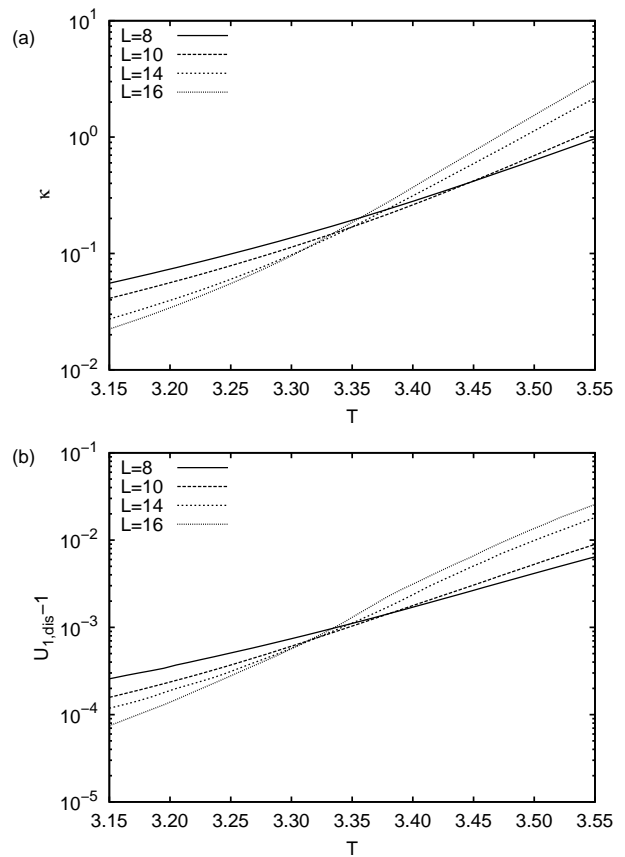}
\caption{\label{fig11} FSS in the RFIM using the path $l_\Gamma$. (a) $\kappa$ 
versus $T$ for various $L$. Note the intersection point, which yields an 
estimate of $\tc$. (b) The disconnected cumulant, \eq{eq:u1dis}, versus $T$ for 
various $L$. The intersections again yield $\tc$. For clarity, the cumulant 
curves are shifted by unity, and a logarithmic vertical scale is used.}
\end{center}
\end{figure}
%% END FIGURE

We now propose one additional method to locate the critical temperature. To this 
end, recall the FSS expressions $\Delta F_L \propto L^\theta$ and $\chi_{\rm 
con} \propto L^{\gamma/\nu}$. Since $\theta=\gamma/\nu$, the ratio $\kappa = 
\chi_{\rm con}/\Delta F_L$ becomes $L$-independent at criticality. One can thus 
locate $\tc$ by plotting $\kappa$ versus $T$ for various system sizes, and look 
for intersection points. This approach has the advantage that the critical 
exponents themselves need not be provided. The connected susceptibility is 
obtained from the individual distributions $\plmi$ using $\chi_{\rm con} = 
(L^d/K) \sum_{i=1}^K (\avg{m^2}_i - \avg{|m|}_i^2)$. In \fig{fig11}(a), we plot 
$\kappa$ versus $T$ for various $L$. The data indeed intersect, providing 
important confirmation that the barrier scales with the same exponent as the 
connected susceptibility at criticality. For completeness, we show in 
\fig{fig11}(b) the disconnected cumulant $\Udis$ versus $T$ for various $L$ (now 
obtained using the path $l_\Gamma$). The curves also intersect, and do so 
remarkably close to the intersections of $\kappa$. Based on \fig{fig11}, we 
(VFB) report $\tcv \approx 3.315 \pm 0.050$, where the error reflects the 
scatter in the various intersection points (here the data of the smallest system 
$L=8$ was ignored).

%% BEGIN FIGURE
\begin{figure}
\begin{center}
\includegraphics[width=0.95\columnwidth]{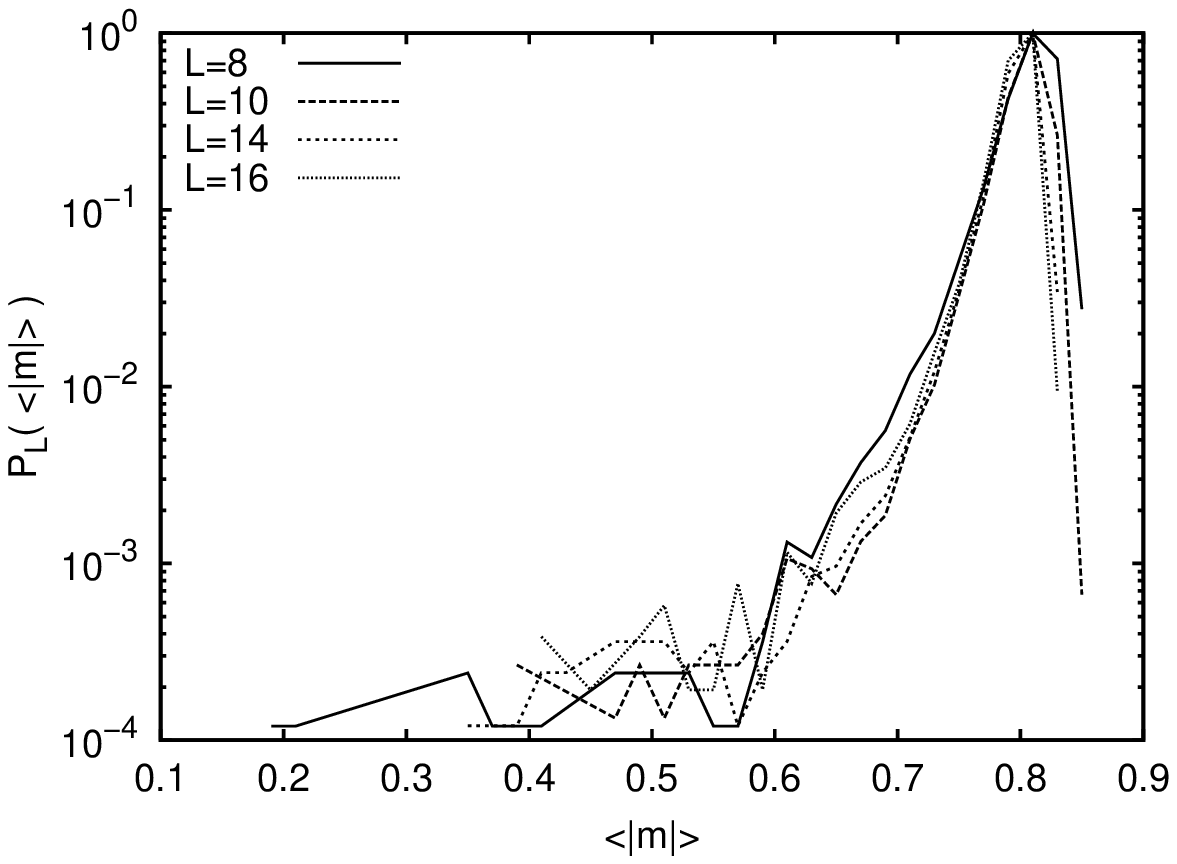}
\caption{\label{fig:calp} $\calp$ for the RFIM at $T=\tcv$, using the path 
$l_\Gamma$, and for various system sizes.}
\end{center}
\end{figure}
%% END FIGURE

We now turn to the distribution $\calp$, defined as the probability of a 
particular random field sample yielding a magnetization $\avg{|m|}$. At 
criticality, we anticipate scaling of this distribution, conform \eq{eq:prf}. We 
have explicitly measured $\calp$ by accumulating a histogram of $\avg{|m|}$ 
values at $T=\tcv$ using the path $l_\Gamma$. The resulting distributions are 
shown in \fig{fig:calp}. The salient features are a sharp peak, and a long tail 
extending to lower values. The fact that $\calp$ features a sharp peak is 
consistent with $\Udis$ being close to unity at criticality. Since 
$\beta \sim 0$ in the RFIM, the scaling variable $x = L^{-\beta/\nu} \avg{|m|}$ 
is identical to $\avg{|m|}$ itself, and so the \ahum{raw} distributions for 
different $L$ should already overlap with each other. Within numerical precision 
this is confirmed, but it is clear that the data in \fig{fig:calp} do not allow 
for any meaningful estimate of $\beta/\nu$.

The point that we wish to make, however, is a different one. The fact that 
$\calp$ features a long tail means that occasionally a distribution $\plmi$ is 
observed with a significantly lower magnetization. Since the scaling form, 
\eq{eq:prf}, implies that $\calp$ retains its shape irrespective of $L$, the 
fraction of these distributions does not vanish in the thermodynamic limit. It 
is conceivable that distributions from the \ahum{tail} of $\calp$ are also 
shaped differently. For instance, consider again the histogram $H(U_1)$ at $\tc$ 
(\fig{fig9}(b)). The histograms peak at $U_1=1$, so most distributions $\plmi$ 
are bimodal. However, $H(U_1)$ also features a tail, so distributions with 
profoundly different shapes, although rare, do occur. In particular, the tail in 
$H(U_1)$ allows for three-peaked distributions to be present (for which 
$U_1=3/2$). Indeed, such distributions are observed, and have been interpreted 
to signify first-order transitions \cite{newman}, or new phases \cite{alvarez}. 
Our point is that the long tail of $\calp$ {\it and} its scale invariance at 
$\tc$ (implied by modified hyperscaling) also allows for the presence of 
three-peaked distributions (without having to assume a first-order transition, 
or the emergence of a new phase).

\section{Widom-Rowlinson model with quenched obstacles}

It was argued by de~Gennes that a binary mixture undergoing phase separation 
inside a random network of quenched obstacles belongs to the universality class 
of the RFIM \cite{39}. The argument is expected to hold when the obstacles 
display a preferred affinity to one of the phases. In case there is no such 
preference, the argument does not apply \cite{sanctis, vink_review}. Previous 
simulations \cite{6} have already produced evidence in favor of de~Gennes' 
argument. To provide further confirmation, in particular to test the scaling of 
the free energy barrier (\eq{eq7}), we consider in this section the 
Widom-Rowlinson binary mixture (WRM) \cite{WidomRowlinson}. The model consists 
of unit diameter spheres, species $A$ or $B$, which may overlap freely except 
for a hard-core repulsion between $A$ and $B$ particles. The model is 
investigated in the grand-canonical ensemble, where the relevant thermodynamic 
parameters are the fugacities, $z_A$ and $z_B$, of the respective species.

At high fugacities, the WRM can be in two phases: a phase rich in $A$ particles 
(the $A$-phase) when $z_A>z_B$, and a phase rich in $B$ particles (the 
$B$-phase) when $z_B>z_A$. Due to the model's symmetry under the exchange of $A$ 
and $B$ particles, the phase transition occurs at $z_A=z_B$. Hence, in line with 
the Ising model, a symmetry path $l_S$ for the WRM can be defined as $l_S: \, 
z_A=z_B$. The transition line ends in an Ising critical point, at fugacity 
$z_A=z_B=\zcp$, below which mixed states appear \cite{WRCP1, WRCP2}. Note that 
the phase transition in the WRM can also be considered a liquid-gas transition. 
By integrating out the $B$ particles, the WRM maps onto a single component 
fluid, interacting via a short-ranged attractive potential 
\cite{WidomRowlinson}. The fugacity $z_B$ then plays the role of inverse 
temperature, the $A$-phase corresponds to the liquid (characterized by a high 
particle density), and the $B$-phase to a gas (low particle density).

\subsection{pure mixture}

%% BEGIN FIGURE
\begin{figure}
\begin{center}
\includegraphics[width=0.95\columnwidth]{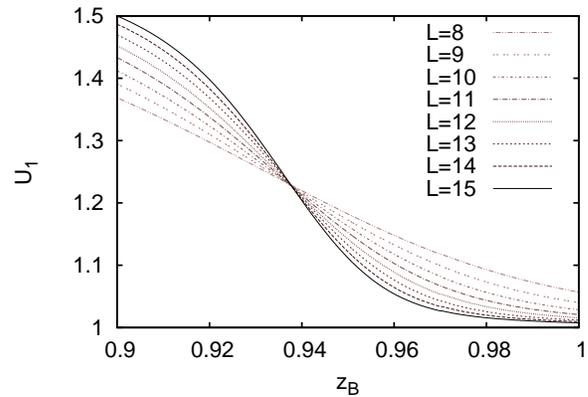}
\caption{\label{fig:WRnodisorder:U1} Cumulant plot for the WRM without quenched 
disorder. Plotted is $U_1$ versus $z_B$ for different $L$. The intersection 
yields $\zcp=0.9377(4)$ and $U_1^\star = 1.228(5)$.}
\end{center}
\end{figure}
%% END FIGURE

%% BEGIN FIGURE
\begin{figure}
\begin{center}
\includegraphics[width=0.95\columnwidth]{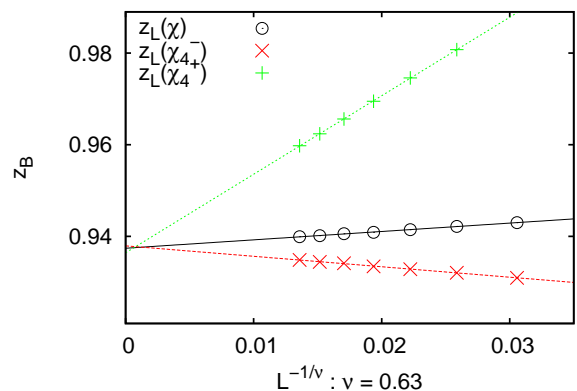}
\caption{\label{fig:WRnodisorder:fss} Extrapolation of $z_L(x)$, $x \in \{ \chi, 
\dchi^-, \dchi^+ \}$, according to \eq{eq:WR:fss} for the WRM with quenched 
disorder. From the linear fits $\zcp = 0.9376(5)$ is obtained.}
\end{center}
\end{figure}
%% END FIGURE

We first consider the pure WRM, i.e.~without quenched obstacles. We simulate 
using cubic boxes with periodic boundary conditions (see Appendix~\ref{app:WR}). 
The analogue of the Ising model OPD is the distribution $P_L(\rho_A)$, defined 
as the probability for a system of lateral extension $L$ to contain $N_A = 
\rho_A L^d$ particles of species $A$. Since we are ultimately interested in 
locating the critical point, only OPDs lying on the symmetry path $l_S$ are 
considered in this section, which leaves $z_B$ as the single free parameter. 
Note that we could also have defined the OPD as $P_L(\rho_A-\rho_B)$, thereby 
directly exploiting the symmetry of the WRM. However, most fluids lack such an 
obvious symmetry, and by using $P_L(\rho_A)$ we ensure that our analysis remains 
generally applicable.

Above the critical fugacity, $z_B>\zcp$, $P_L(\rho_A)$ is bimodal: the peak at 
low (high) density corresponds to the gas (liquid) phase. When $z_B<\zcp$, the 
OPD features a single peak, corresponding to a mixed state. The analogue of the 
magnetization is defined as $ m \equiv \rho_A - \avg{\rho_A}$, which is readily 
substituted in \eq{eq:sus} to yield the order parameter and susceptibility. 
Additionally, we define a \ahum{generalized} susceptibility \cite{orkoulas, 
locus}
\bea
\begin{split}
 \dchi \equiv L^{3d} \left( \avg{m^4} - 4 \avg{|m|} \avg{|m|^3} + 12 \avg{m^2} \avg{|m|}^2 \right. \\
   \left. - 3 \avg{m^2}^2 - 6 \avg{|m|}^4 \right). \hspace{2cm}
\end{split}
\eea
The most straightforward method to locate the critical point is from 
intersections of the Binder cumulant for different $L$. For the pure WRM, we 
find that a sharp intersection of $U_1$ can be found easily 
(\fig{fig:WRnodisorder:U1}). Another method to locate the critical fugacity is 
via the extrapolation of the finite-size extrema of $\chi$ and $\dchi$. In a 
finite system of size $L$, $\chi$ reaches a maximum at fugacity $z_L(\chi)$, 
which is shifted from $\zcp$ as \cite{locus}
\beq \label{eq:WR:fss}
  \zcp - z_L(\chi) \propto L^{-1/\nu},
\eeq
with $\nu$ the correlation length critical exponent. In addition, $\dchi$ 
reaches a minimum and maximum, at respective fugacities $z_L(\dchi^-)$ and 
$z_L(\dchi^-)$, which are also shifted according to \eq{eq:WR:fss}. Hence, 
plotting $z_L(\chi)$, $z_L(\dchi^-)$, and $z_L(\dchi^+)$ versus $L^{-1/\nu}$, 
and then linearly extrapolating to $L \to \infty$, three additional estimates of 
$\zcp$ are obtained. For this extrapolation, hyperscaling is not required, but 
$\nu$ needs to be provided. In principle, $\nu$ can also be taken as a fit 
parameter, but this requires data of extremely high quality. For the pure WRM, 
which belongs to the Ising universality class, $\nu$ is known (cf.~\tab{tab1}). 
In \fig{fig:WRnodisorder:fss}, the extrapolation is demonstrated; the resulting 
estimates of $\zcp$ are similar and agree with the cumulant intersections. 
Combining all results, we obtain $\zcp=0.9377(5)$, where the error reflects the 
scatter between the individual estimates. This value is in good agreement with 
previous results \cite{WRCP1, WRCP2}.

\subsection{mixture with quenched obstacles}

%% BEGIN FIGURE
\begin{figure}
\begin{center}
\includegraphics[width=0.95\columnwidth]{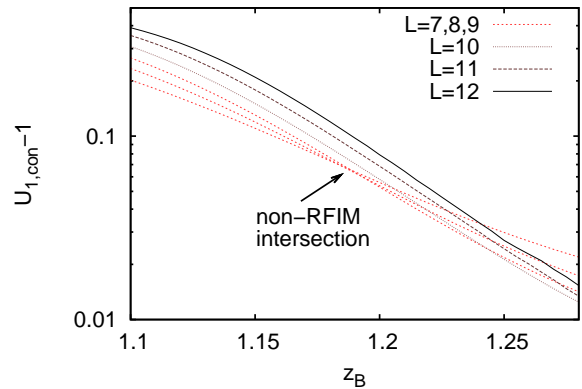}
\caption{\label{fig:WR:U1} Cumulant plot for the WRM with quenched obstacles. 
Plotted is $\Ucon$ (\eq{eq:u1rf}) versus $z_B$ for different $L$. The 
intersection of system sizes $L<10$ is attributed to a crossover effect from 
Ising universality to RFIM universality. Curves for $L\geq 10$ no longer 
intersect at this point indicating that for these system sizes the crossover to 
RFIM universality has largely completed.}
\end{center}
\end{figure}
%% END FIGURE

%% BEGIN FIGURE
\begin{figure}
\begin{center}
\includegraphics[width=0.95\columnwidth]{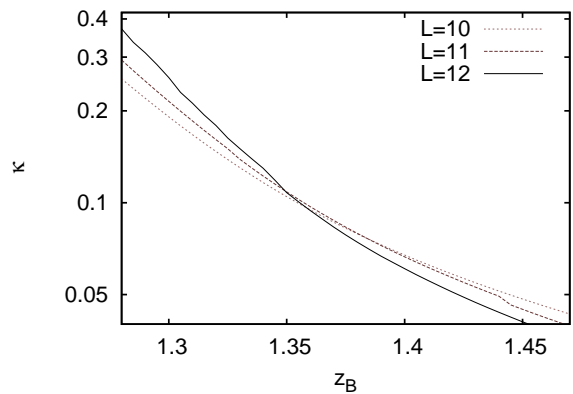}
\caption{\label{fig:WR:dFoverChi} Variation of $\kappa$ with $z_B$ for various 
$L$ for the WRM with quenched obstacles (analogue of \fig{fig11}(a)). The 
intersection of the curves indicates a critical point around $z_B \sim 1.35$.}
\end{center}
\end{figure}
%% END FIGURE

%% BEGIN FIGURE
\begin{figure}
\begin{center}
\includegraphics[width=0.95\columnwidth]{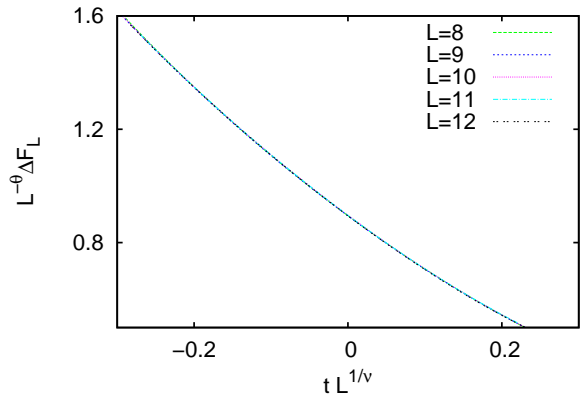}
\caption{\label{fig:WR:BarrierScaling} Scaling plot of the free energy barrier, 
according to \eq{eq:gx}, for the WRM with quenched obstacles (analogue of 
\fig{fig10}(b)).}
\end{center}
\end{figure}
%% END FIGURE

%% BEGIN FIGURE
\begin{figure}
\begin{center}
\includegraphics[width=0.95\columnwidth]{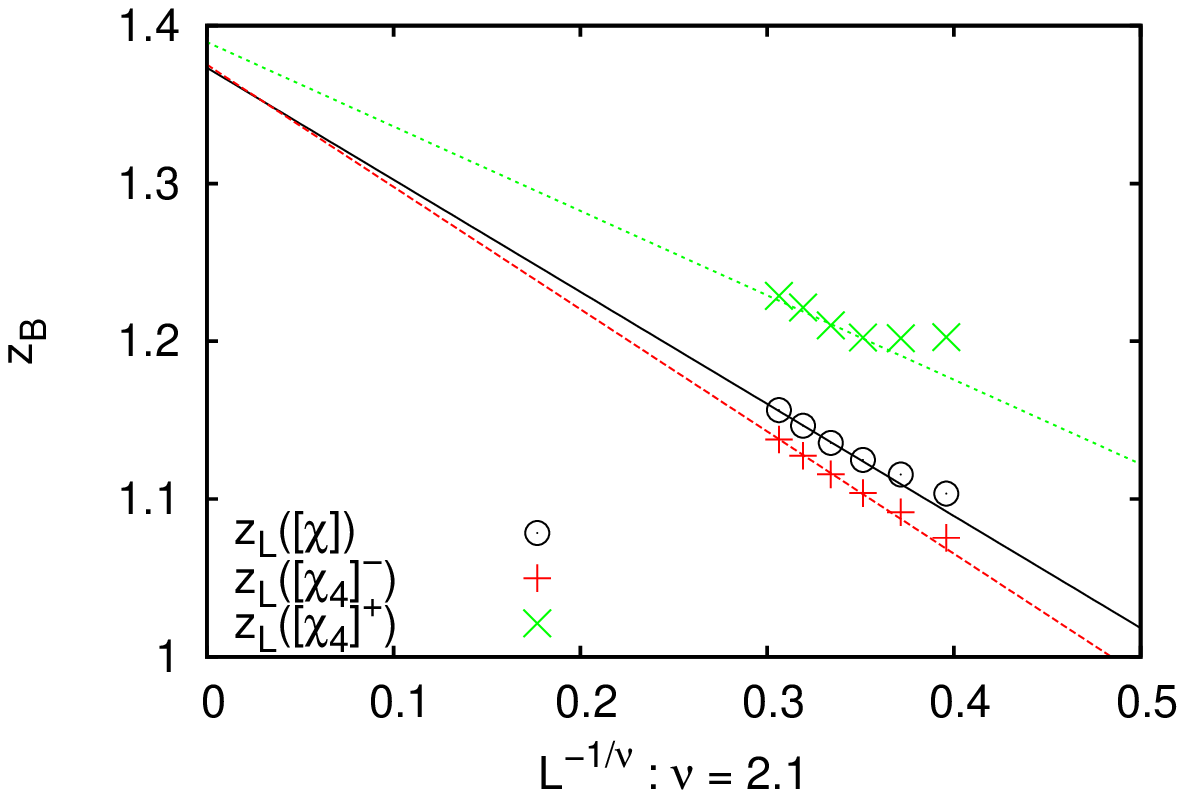}
\caption{\label{fig:WR:fss} Extrapolation of $z_L(x)$, $x \in \{ [\chi], 
[\dchi]^-, [\dchi]^+ \}$, via \eq{eq:WR:fss} for the WRM with quenched disorder. 
The plot uses $\nu=2.1$ (as obtained in \fig{fig:WR:BarrierScaling}), and 
predicts $\zcp=1.37(2)$ in the limit $L \to \infty$.}
\end{center}
\end{figure}
%% END FIGURE

We now consider the WRM model with quenched obstacles. We use spherical 
obstacles, species $X$ and $Y$, having the same diameter as the (mobile) $A$ and 
$B$ particles. The total number of $X$ and $Y$ obstacles equals $N_X = N_Y = 
\rho_Q \times L^d$, rounded up or down at random to the next integer. The 
obstacles are distributed randomly at the start of the simulation, irrespective 
of overlap, after which they remain quenched: this defines one disorder 
realization~$i$. Next, $A$ and $B$ particles are introduced, and grand canonical 
MC is used to construct $P_{L,i}(\rho_A)$ for that disorder realization (see 
Appendix). The $A$-particles ($B$-particles) have a hard-core interaction with 
$X$-obstacles ($Y$-particles) but do not interact with $Y$-obstacles 
($X$-obstacles). The original motivation for this choice was to restore the 
symmetry line $l_S : z_A=z_B$ in the disorder average. However, in what follows, 
we will use the path $l_\Gamma$, whereby $z_A$ is tuned for each realization of 
disorder such that $\partial \avg{\rho_A} / \partial \log z_A$ is maximized.

We still need to specify the obstacle concentration $\rho_Q$. For a noticeable 
random field effect, the thermal correlation length $\xi$ should be large 
compared to the typical distance $\xi_Q$ between obstacles. Following the FSS 
\ahum{Ansatz} $\xi \propto L$, this implies $L \gg \xi_Q$. If $L$ is too small, 
crossover scaling is observed (in this case from pure Ising to random field 
Ising \cite{6}). From these considerations, choosing a high value of $\rho_Q$ 
seems optimal. The disadvantage is that also $\zcp$ will then be very high, 
which makes the grand canonical MC approach inefficient due to a high particle 
density. Clearly, a compromise needs to be made: we use $\rho_Q=0.02$. This 
value is small compared to typical density of the mobile species, e.g.~$\rho_A 
=\rho_B \approx 0.38$ at criticality in the pure WRM, and certainly is below the 
percolation threshold; we thus remain in the limit of weak random fields. For 
the chosen obstacle density, crossover effects are still strong in small 
systems. This can be inferred from \fig{fig:WR:U1}, where the connected cumulant 
$\Ucon$ versus $z_B$ for various $L$ is plotted. The curves for $L<10$ reveal an 
intersection point: this would be consistent with a conventional critical point 
featuring standard hyperscaling. However, for $L \geq 10$, the intersection has 
vanished, indicating that the crossover has largely completed. In what follows, 
we therefore discard the data for $L<10$ in some of the analysis.

Investigations involving disconnected quantities require enormous simulational 
effort to generate data of sufficient quality (see \fig{fig:ra} in 
Appendix~\ref{sec:ra}). For the WRM, an analysis of $\Udis$ along the lines of 
\fig{fig11}(b) was not feasible. We therefore focus on the free energy barrier. 
We evaluate the distributions $P_{L,i}(\rho_A)$ along the path $l_\Gamma$, and 
for each distribution, we \ahum{read-off} the barrier, which is then averaged 
over the samples to obtain $\Delta F_L$. We first consider the variation of 
$\kappa$ versus $z_B$ for different $L$, i.e.~the analogue of \fig{fig11}(a). 
This data is shown in \fig{fig:WR:dFoverChi}; from the intersection we conclude 
that the critical fugacity is around $z_B \sim 1.35$. To get the critical 
exponents, we consider the scaling of the free energy barrier. Assuming RFIM 
universality, the variation of $\Delta F_L$ with $z_B$ should follow \eq{eq:gx}, 
where now $t \equiv (\zcp-z_B)/z_B$. In the vicinity of $z_B \sim 1.35$, i.e.~as 
indicated by \fig{fig:WR:dFoverChi}, we indeed find that a collapse of the 
curves can be realized for $\theta=1.32$, $\nu=2.1$, and $\zcp=1.37$ 
(\fig{fig:WR:BarrierScaling}). As a consistency check, we attempt to obtain 
$\zcp$ from the extrapolation of the extrema of the susceptibilities using 
\eq{eq:WR:fss}. The observables $\chi$ and $\dchi$ of the pure model are now 
replaced by their disorder-averaged counterparts $[\chi]$ and $[\dchi]$, and 
$\nu=2.1$, i.e.~the estimate from \fig{fig:WR:BarrierScaling}, is used. The 
extrapolation works reasonably well (\fig{fig:WR:fss}) and for the critical 
fugacity we obtain the same estimate as before:~$\zcp=1.37(2)$.

\section{Summary}

Modified hyperscaling, \eq{eq2}, which is believed to describe systems belonging 
to the universality class of the RFIM, gives rise to rather unusual finite size 
effects at critical points: neither the order parameter distribution, nor the 
free energy barrier $\Delta F_L$ of interface formation, are scale invariant. As 
a result, \ahum{standard} techniques to locate critical points, such as the 
\ahum{cumulant intersection method} \cite{15}, or the Lee-Kosterlitz method 
\cite{lee}, break down. However, by carefully considering the consequences of 
\eq{eq2}, alternative techniques to derive $\tc$ in random field systems can be 
derived. In this paper, we have proposed two such techniques. The first is based 
on the order parameter fluctuations between disorder samples: modified 
hyperscaling predicts that these are scale invariant at $\tc$. This property can 
be used to locate $\tc$ by measuring the {\it disconnected} cumulant $\Udis$ 
(\eq{eq:u1dis}) versus temperature for various system sizes: at $\tc$, curves 
for different $L$ intersect. Indeed, simulation data of the RFIM confirm the 
scaling of $\Udis$ (\fig{fig6}(b) and \fig{fig11}(b)). In contrast to 
conventional critical points, there is no intersection of the {\it connected} 
cumulant $\Ucon$ (\eq{eq:u1rf}) in the RFIM at $\tc$. However, in small systems, 
there may be crossover effects. In this case, an apparent intersection in 
$\Ucon$ is observed, at $T>\tc$, but it vanishes in larger systems; such was the 
case for the WRM (\fig{fig:WR:U1}).

The practical disadvantage of measuring $\Udis$ is that many disorder samples 
must be averaged over if meaningful results are to be obtained. Particularly for 
more complex systems, such as {\it off-lattice} fluids, an economic alternative 
is to consider the free energy barrier $\Delta F_L$ of interface formation. Due 
to modified hyperscaling, the barrier diverges $\Delta F_L \propto L^\theta$ at 
$\tc$, with $\theta$ the violation of hyperscaling exponent. The consequences of 
this divergence are easily detected in simulations, as was demonstrated for the 
RFIM (\fig{fig10} and \fig{fig11}(a)), and the WRM (\fig{fig:WR:dFoverChi} and 
\fig{fig:WR:BarrierScaling}). In case of the RFIM, the estimate of $\tc$ 
obtained from the scaling of the barrier was fully consistent with that obtained 
from the intersections of $\Udis$ (\fig{fig11}). Our results for the WRM provide 
further confirmation that fluids with quenched disorder indeed belong to the 
universality class of the RFIM, consistent with the conjecture of de~Gennes 
\cite{39}.

We have also commented on the variations in shape of the order parameter 
distribution between samples. There is some question as to whether distributions 
with three peaks signify first-order transitions \cite{newman}, or the emergence 
of new phases \cite{alvarez}. Our view is that modified hyperscaling also allows 
for these shape variations. While our data indicate that at $\tc$, and using the 
path $l_\Gamma$, the majority of distributions is bimodal, a fraction of 
distributions with different shape is not ruled out (\fig{fig9}(b) and 
\fig{fig:calp}). 

Finally, we remind the reader that the divergence of the free energy barrier at 
$\tc$ will also influence the dynamics. Taking the RFIM with single spin-flip 
dynamics as example, it follows that the largest relaxation time in a finite 
system at criticality is given by an Arrhenius' type formula
\begin{equation}\label{binder}
 \ln \tau \propto L^\theta \quad (T=\tc).
\end{equation}
This is in contrast to the pure model, where the relaxation time (not its 
logarithm) scales $\tau \propto L^z$ , with $z$ the \ahum{dynamical critical 
exponent}. Such a power law for the logarithm of the relaxation time is the 
hallmark of \ahum{activated critical dynamics}. In fact, if we are somewhat 
above $\tc$, but the system size is still less than the correlation length, 
$L<\xi$, \eq{binder} still holds! As $L>\xi$, the system size in \eq{binder} 
gets replaced by $\xi$, and we recover \eq{eq5}, as proposed by Villain 
\cite{26} and Fisher \cite{27}. A direct study of the dynamics of a kinetic 
version of the RFIM would be illuminating, but goes beyond the scope of the 
present paper.

\acknowledgments

This work was supported by the {\it Deutsche Forschungsgemeinschaft}
(Emmy Noether program:~VI~483/1-1).

\appendix*

\section{Simulation Details}

\subsection{Wang-Landau sampling}

The simulations of the Ising and RFIM were performed using Wang-Landau (WL) 
sampling \cite{wls}. The OPD is written as
\begin{equation}\label{eq:wl}
 P_{L,i} (S) \propto g_{L,i} (S) e^{H S / k_B T}, 
\end{equation} 
with $S = L^d m$ the total instantaneous magnetization, and $g_{L,i}(S)$ some 
generalized density of states (DOS). Note that the DOS depends on system size 
$L$, temperature $T$, and random field sample $i$, but not on the external field 
$H$ (for the pure Ising model, there is no dependence on $i$ either, of course). 
At the start of each simulation, we generate a sample of random fields $i$. We 
then perform single spin-flips, whereby one of the spins is chosen at random, 
and its orientation reversed. Let the total magnetization and energy at the 
start of each spin-flip be given by $S_0$ and $E_0$, respectively, and afterward 
by $S_1$ and $E_1$; each spin-flip is then accepted with probability
\begin{equation}\label{eq:pacc}
\begin{split}
 a(S_0,&E_0 \to S_1,E_1) = \\ & \min \left[1, 
 \frac{ g_{L,i}(S_0) }{ g_{L,i}(S_1) } e^{-(E_1 - E_0) / k_B T} \right].
\end{split}
\end{equation}
Note that the energy above refers to the configurational part of the Hamiltonian 
only, i.e.~the nearest-neighbor interaction and the coupling to the random 
field, but not the coupling to the external field.

The DOS is {\it a-priori} unknown, and is initially set to unity $g_{L,i}(S)=1$. 
After each attempted spin-flip, one \ahum{updates} the DOS $g_{L,i}(S) \to f 
\times g_{L,i}(S)$, with $S$ the magnetization of the system after the attempted 
spin-flip. The update is performed irrespective of whether the spin-flip was 
accepted; the initial modification factor $f = e \approx 2.72$. We also update a 
histogram $h(S) \to h(S)+1$, counting how often a state with magnetization $S$ 
was visited. This procedure is repeated until $h(S)$ has become sufficiently 
flat, which completes one WL iteration. We use the criterion $(h_{\rm max} - 
h_{\rm min}) / (h_{\rm max} + h_{\rm min}) < 10^{-5}$, with $h_{\rm min}$ and 
$h_{\rm max}$ the smallest and largest entries in $h(S)$, respectively. After 
the first WL iteration, the modification factor is reduced $f \to f^{1/2}$, the 
histogram $h(S)$ is reset to zero, and the procedure is repeated. WL iterations 
are continued until $f$ has become small such that changes to the DOS become 
negligible. For each DOS, we typically performed 150--250~WL iterations. Once 
the DOS is known, the OPD can be calculated for arbitrary values of $H$ using 
\eq{eq:wl}.

\subsection{Importance of disorder averaging}

\label{sec:ra}
%% BEGIN FIGURE
\begin{figure}
\begin{center}
\includegraphics[width=0.95\columnwidth]{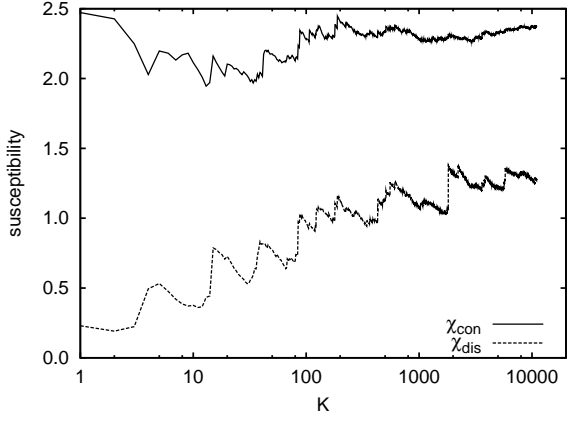}
\caption{\label{fig:ra} \ahum{Running averages} of the connected and the 
disconnected susceptibility versus the number of random field samples $K$. The 
data were obtained for the RFIM using the symmetry path $l_S$, $L=14$, and 
$T=\tcv$. Note the logarithmic horizontal scale.}
\end{center}
\end{figure}
%% END FIGURE

To accurately determine disorder averages, the OPD $\plmi$ is measured $i=1, 
\ldots ,K$ times. In particular disconnected quantities require a large number 
of disorder samples if meaningful results in the critical regime are to be 
obtained. \fig{fig:ra} shows a typical \ahum{running average} of $\chi_{\rm 
con}$ and $\chi_{\rm dis}$ versus $K$. While $\chi_{\rm con}$ saturates to a 
plateau already after 1000~samples, the convergence of $\chi_{\rm dis}$ is 
noticeably slower. The data of \fig{fig:ra} indicate that $K$ should exceed 
several thousands at least. Away from the critical point, $\chi_{\rm dis}$ is no 
longer divergent, and here we expect that lower values of $K$ will also suffice.

\subsection{Histogram reweighting in temperature}
\label{app:histogramreweighting}

A key ingredient in this work is the use of histogram reweighting in the 
temperature-like variable. We perform our simulations at only a few distinct 
temperatures, and extrapolate to other values using histogram reweighting 
\cite{hrw}. This requires that the joint two-dimensional probability 
distribution $P_{L,i}(S,E)$, of the magnetization $S$ and energy $E$, is known. 
Again, as in \eq{eq:pacc}, $E$ refers to the configurational part of the 
Hamiltonian only. If $P_{L,i}(S,E)$ is measured for $T=T_0$ and $H=H_0$, it can 
be extrapolated to other values using a generalization of \eq{eq:wl}
\begin{equation}
 P_{L,i} (S,E) |_{T_1,H_1} \propto P_{L,i} (S,E) |_{T_0,H_0} 
 \times e^{\delta h S -\delta\beta E},
\end{equation}
with $\delta h = (H_1-H_0)/k_B T_1$ and $\delta\beta = 1/k_B T_1 - 1/k_B T_0$. 
The practical problem is that two-dimensional histograms require considerable 
disk space, which in the case of quenched disorder is multiplied by a factor 
$K$. Fortunately, an excellent approximation can be used to drastically reduce 
storage requirements \cite{6}. Without loss of generality, we write
\begin{equation}\label{eq:compact}
 P_{L,i} (S,E) = P_{L,i}(S) \times g_{L,i}^{(S)} (E),
\end{equation}
where $g_{L,i}^{(S)} (E)$ is the probability distribution of the energy measured 
at states with the same magnetization $S$. The approximation is to assume that 
$g_{L,i}^{(S)} (E)$ is Gaussian, and so is fully specified by its first two 
moments. For each random field sample, the two-dimensional histogram of 
\eq{eq:compact} then requires only $P_{L,i}(S)$ to be stored, plus the 
\ahum{functions} $\avg{E}_{L,i}(S)$ and $\avg{E^2}_{L,i}(S)$.

\subsection{Alternative method to measure the barrier}
\label{app:alt}

%% BEGIN FIGURE
\begin{figure}
\begin{center}
\includegraphics[width=0.95\columnwidth]{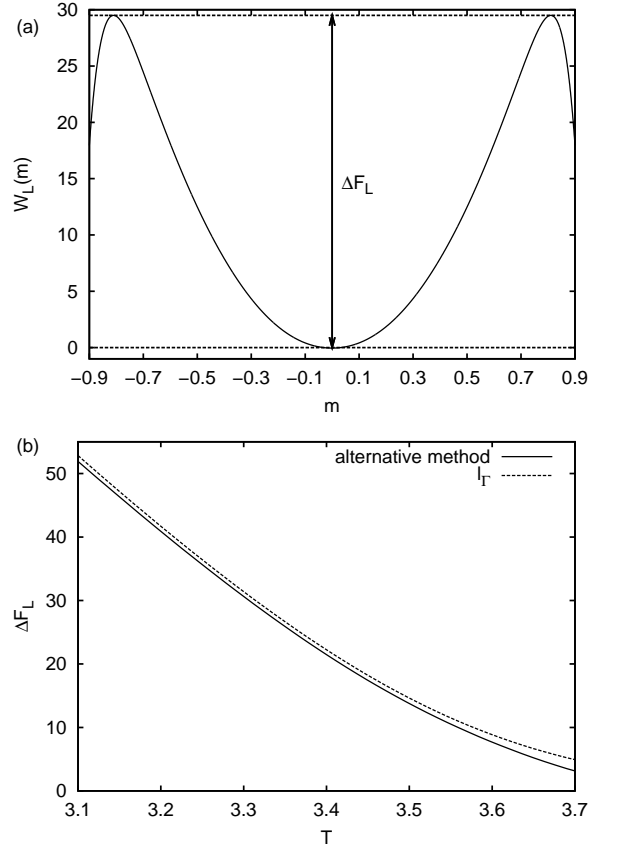}
\caption{\label{fig:bar_alt} Demonstration of an alternative method to extract 
the free energy barrier $\Delta F_L$. The data in this figure were obtained for 
the RFIM using $L=14$. (a) The quenched-averaged free energy distribution 
$W_L(m)$ constructed with the recursion relation of \eq{eq:rec} at $T=\tcv$; a 
free energy barrier $\Delta F_L$ can be meaningfully extracted. (b) The 
corresponding variation of $\Delta F_L$ versus $T$, compared to the 
\ahum{original} method, where $\Delta F_L$ is averaged over individual samples 
using the path $l_\Gamma$.}
\end{center}
\end{figure}
%% END FIGURE

It is also possible to measure the quenched-averaged free energy barrier $\Delta 
F_L$ using the same external field for all samples. To be concrete, consider the 
OPD $P_{L,i}(S)$ of the RFIM obtained at fixed $H=0$, i.e.~the symmetry path, 
with total magnetization $S = -L^d,-L^d+2,\ldots,L^d$ and $i=1, \ldots ,K$. We 
define the quenched-averaged free energy difference between \ahum{adjacent} 
states as
\begin{equation}
 \Delta W_L(S-2,S) = \frac{1}{K} \sum_{i=1}^K \ln \left( 
 \frac{ P_{L,i}(S) }{ P_{L,i}(S-2) } \right),
\end{equation}
which can be used to construct a total free energy $W_L(S)$ by means of 
recursion
\begin{equation}\label{eq:rec}
\begin{split}
 &W_L(-L^d) \equiv 0, \\
 &W_L(S) = W_L(S-2) + \Delta W_L(S-2,S). 
\end{split}
\end{equation}
\fig{fig:bar_alt}(a) shows the typical shape of the free energy obtained in this 
way for the RFIM. The distribution is bimodal, and a free energy barrier $\Delta 
F_L$ can be meaningfully \ahum{read-off}. As it turns out, this barrier is very 
similar to that obtained by averaging over individual samples, i.e.~as was done 
in \fig{fig10}(a) using the path $l_\Gamma$; a comparison is provided in 
\fig{fig:bar_alt}(b). In fact, if one uses $W_L(S)$ to perform the scaling 
analysis of \fig{fig10}(b), excellent data collapses are also realized.

An analysis in terms of $W_L(S)$ is numerically convenient because 
extrapolations in the field variable $H$ can be performed after the quenched 
average has been taken
\begin{equation}
 \left. W_L(S) \right|_{H_1} = 
 \left. W_L(S) \right|_{H_0} + (H_1 - H_0) S / k_B T.
\end{equation}
This is particularly useful for fluids, where the critical field (chemical 
potential) is generally not known beforehand. However, it is not obvious what 
the peak positions and widths in $W_L(S)$ correspond to. Based on our previous 
work \cite{6}, cumulants of $e^{W_L(S)}$ do intersect at $\tc$, but (in 
hindsight) we believe it is safer to perform the cumulant analysis using the 
individual OPDs (as was done in this work).

\subsection{Simulating the Widom-Rowlinson model}
\label{app:WR}

We measure $P_{L,i}(\rho_A)$ using grand canonical MC and successive umbrella 
sampling (SUS) \cite{SUS}. The simulations are performed in a periodic 3D cube 
of volume $V=L^3$. At the start of each simulation, a disorder realization $i$ 
is generated by distributing the obstacles $X$ and $Y$ randomly in the cube, 
i.e.~the obstacles are allowed to overlap. We then perform grand canonical MC 
moves consisting of the insertion and removal of single $A$ and $B$ particles.
In SUS, the full density range of interest is split into overlapping windows 
$W_k$. In the first window, $N_A$ is allowed to fluctuate between 0 and 1, in 
the second window between 1 and 2, or, more generally, in the $k$-th window $W_k
: N_A \in \{k,k+1\}$. There is no restriction on the number of $B$ particles: 
$N_B$ thus fluctuates freely in each window.

For $N_A=0$ the $B$-particles are an ideal gas in the volume allowed by the 
quenched $Y$-particles so an initial state for $W_0$ is easily constructed: we 
draw a number $N$ from a Poissonian distribution $P(N) = e^{-z_B V} (z_B V)^N / 
N!$, and randomly insert this number of $B$-particles into the system, 
discarding all $B$-particles that overlap with $Y$-obstacles. As starting state 
for the subsequent windows $W_k \, (k>0)$, we take the last state of the window 
$W_{k-1}$ preceding it, and equilibrate this state briefly for $\sim 10^5$ MC 
steps within the bounds of the new window. This works well in practice because 
the windows are small.

The production run of each window $W_k$ is performed using $\sim 10^7$ MC steps. 
Each step first selects a species, $x \in \{A,B\}$, with equal probability. 
Then, with equal probability, the insertion or removal of a particle of species 
$x$ is attempted. In case of removal, a particle of species $x$ is picked at 
random and removed from the system; the resulting new state is accepted with 
probability
\beq
 a( N_x \to N_x-1) =
 \text{min} \left[ 1, \frac{N_x}{z_x V} f_k^{-1} \right].
\eeq
The factor $f_k$ is one when $x=B$; for $x=A$, it will be specified later. 
In case insertion is chosen, a new particle of species $x$ is placed at a random 
location; the resulting new state is accepted with probability
\beq
 a( N_x \to N_x +1)= 
 \text{min} \left[ 1, \frac{z_x V}{N_x+1} f_k \right].
\eeq
States with hard-core overlaps and states where $N_A$ is outside the window 
bounds are always rejected, irrespective of the accept probabilities.

While simulating in window $W_k$, we keep track of two counters, $C_k^-$ and 
$C_k^+$. These count, respectively, how often the state with $N_A=k$ and 
$N_A=k+1$ was visited. From these counters, we construct the relative 
probability of these states via
\beq
 \label{eq:SUSratios}
 \frac{P_{L,i}(k+1)}{P_{L,i}(k)} \approx \frac{C_k^+}{C_k^-} f_k^{-1}.
\eeq
Having at hand this ratio for all windows $W_k$, the full distribution is 
constructed recursively
\beq
 P_{L,i}(N_A) \propto
 \prod_{k=0}^{N_A-1} \frac{P_{L,i}(k+1)}{P_{L,i}(k)}
 \approx \prod_{k=0}^{N_A-1} \frac{C_k^+}{C_k^-} f_k^{-1}, 
\eeq
where the proportionality constant follows from normalization. Note that 
$P_{L,i}(N_A)$ above is, of course, fully equivalent to the OPD 
$P_{L,i}(\rho_A)$ that we wish to find.

We now specify the factor $f_k$ for moves involving $A$-particles. Assuming a 
constant number of steps per window, \eq{eq:SUSratios} suggests that optimal 
results are obtained when $f_k$ is chosen such that $C_k^+$ and $C_k^-$ are 
roughly equal, i.e.~$f_k = P_{L,i}(k) / P_{L,i}(k+1)$, which is the sought-for 
result itself. For the first window $k=0$ we use the pure model's optimal 
weight, which can be calculated analytically. For the subsequent windows, we 
linearly extrapolate $P_{L,i}(N_A)$ to calculate $f_k = P_{L,i}(k-1) / 
P_{L,i}(k)$ to be used in that window. In practice, this choice is already quite 
good, and the counts in the upper and lower bin consistently lie within $1\%$ of 
each other.

In view of the huge amount of disorder realizations required, a mechanism that 
allows for histogram reweighting of results obtained at $(z_A, z_B)$ to nearby 
parameters $(\bar z_A, \bar z_B)$ is indispensable. To facilitate this 
reweighting, the joint probability distribution $P_{L,i}(N_A, N_B)$ is stored in 
compact form as described in Appendix~\ref{app:histogramreweighting}; the 
results of that section trivially transfer to the WRM if one identifies $S 
\leftrightarrow N_A$ and $E \leftrightarrow N_B$. For the WRM with quenched 
disorder, the range in $z_B$ over which one can reliably extrapolate is too 
small to cover the full region of interest. In particular, simulation data 
obtained at the fugacity $z_L([\chi])$ of the susceptibility maximum could not 
be extrapolated to the critical fugacity $\zcp$. We therefore created two data 
sets per system size. One set with $K=2000$ disorder realizations at $z_B 
\approx z_L([\chi])$ used for locating the extrema of $\chi$ and $\dchi$ 
(\fig{fig:WR:fss}) and one set with $K=10000$ realizations around $z_B \approx 
1.4$, which is close to $\zcp$, for investigating the free energy barrier 
(Figs.~\ref{fig:WR:dFoverChi} and~\ref{fig:WR:BarrierScaling}).

\end{document}